\documentstyle[12pt]{article}
\makeatletter
\newdimen\normalarrayskip              
\newdimen\minarrayskip                 
\normalarrayskip\baselineskip
\minarrayskip\jot
\newif\ifold             \oldtrue            \def\new{\oldfalse}
\def\arraymode{\ifold\relax\else\displaystyle\fi} 
\def\eqnumphantom{\phantom{(\theequation)}}     
\def\@arrayskip{\ifold\baselineskip\z@\lineskip\z@
     \else
     \baselineskip\minarrayskip\lineskip2\minarrayskip\fi}
\def\@arrayclassz{\ifcase \@lastchclass \@acolampacol \or
\@ampacol \or \or \or \@addamp \or
   \@acolampacol \or \@firstampfalse \@acol \fi
\edef\@preamble{\@preamble
  \ifcase \@chnum
     \hfil$\relax\arraymode\@sharp$\hfil
     \or $\relax\arraymode\@sharp$\hfil
     \or \hfil$\relax\arraymode\@sharp$\fi}}
\def\@array[#1]#2{\setbox\@arstrutbox=\hbox{\vrule
     height\arraystretch \ht\strutbox
     depth\arraystretch \dp\strutbox
     width\z@}\@mkpream{#2}\edef\@preamble{\halign \noexpand\@halignto
\bgroup \tabskip\z@ \@arstrut \@preamble \tabskip\z@ \cr}%
\let\@startpbox\@@startpbox \let\@endpbox\@@endpbox
  \if #1t\vtop \else \if#1b\vbox \else \vcenter \fi\fi
  \bgroup \let\par\relax
  \let\@sharp##\let\protect\relax
  \@arrayskip\@preamble}
\def\eqnarray{\stepcounter{equation}%
              \let\@currentlabel=\theequation
              \global\@eqnswtrue
              \global\@eqcnt\z@
              \tabskip\@centering
              \let\\=\@eqncr
              $$%
 \halign to \displaywidth\bgroup
    \eqnumphantom\@eqnsel\hskip\@centering
    $\displaystyle \tabskip\z@ {##}$%
    &\global\@eqcnt\@ne \hskip 2\arraycolsep
         \hfil$\arraymode{##}$\hfil
    &\global\@eqcnt\tw@ \hskip 2\arraycolsep
         $\displaystyle\tabskip\z@{##}$\hfil
         \tabskip\@centering
    &{##}\tabskip\z@\cr}
\makeatother

\def\beq{\begin{equation}}
\def\eeq{\end{equation}}
\def\bea{\begin{eqnarray}}
\def\eea{\end{eqnarray}}

\def\stackreb#1#2{\mathrel{\mathop{#2}\limits_{#1}}}
\def\theequation{\thesection.\arabic{equation}}  

\begin{document}

\begin{titlepage}
\begin{center}
{{\it P.N.Lebedev Institute preprint} \hfill FIAN/TD-03/92\\
{\it I.E.Tamm Theory Department} \hfill ITEP-M-3/92
\begin{flushright}{February 1992}\end{flushright}
\vspace{0.1in}{\Large \bf Generalized Kontsevich Model}\\{\Large \bf
versus Toda hierarchy}\\{\Large \bf
and discrete matrix models}\\[.8in]
{\large  S.Kharchev, A.Marshakov, A.Mironov}\\
\bigskip {\it  P.N.Lebedev Physical
Institute \\ Leninsky prospect, 53, Moscow, 117 924},
\footnote{E-mail address: theordep@sci.fian.msk.su}\\ \smallskip
\bigskip {\large A.Morozov}\\
 \bigskip {\it Institute of Theoretical and Experimental
Physics,  \\
 Bol.Cheremushkinskaya st., 25, Moscow, 117 259}\footnote{E-mail address:
 morozov@itep.msk.su}}
\end{center}
\bigskip
\bigskip

\newpage
\setcounter{page}2
\centerline{\bf ABSTRACT}
\begin{quotation}

We represent the partition function of the Generalized Kontsevich Model (GKM)
in the form of a Toda lattice $\tau$-function and discuss various
implications of
non-vanishing "negative"- and "zero"-time variables: the appear to modify the
original GKM action by negative-power and logarithmic contributions
respectively. It is shown that so deformed $\tau$-function satisfies the
same string equation as the original one.
In the case of quadratic potential GKM turns out to describe
{\it forced} Toda chain hierarchy and, thus, corresponds to a {\it discrete}
matrix model, with the role of the matrix size played by the zero-time
(at integer positive points). This relation allows one to discuss the
double-scaling continuum limit entirely in terms of GKM, $i.e.$ essentially
in terms of {\it finite}-fold integrals.
\end{quotation}
\end{titlepage}

\newpage
\setcounter{footnote}0
\section{Introduction}

In [1] a new matrix model was introduced which interpolates between all the
non-per\-tur\-ba\-ti\-ve
partition functions of Virasoro $(q,q')$-minimal string models
with  $\displaystyle{c = 1 - {6(q-q')^2\over qq'}}$. The partition function of
GKM depends on
two distinct sets of time-variables: one entering the ``potential"
$\displaystyle{V(X)
=\sum _{p\geq 1} {s_pX^{_p}\over p}}$, and the other one is connected by a
kind of Miwa relation with auxiliary matrix  $M$:

\beq
t_p = {1\over p}[TrM^{-p} + (p-1)s_p].
\eeq
For the particular choice of parameters (potential  $V(X) =
const\cdot X^{q+1})$ it coincides with the description of
$(q,q')$-series\footnote{$E.g$. for $q=2$ -- with Witten's topological gravity
[2], which has
been represented in the form of a matrix model by Kontsevich [3].}.
Interpolation as given by GKM preserves both nice properties of the known
non-perturbative partition functions: if considered as functions of
$t$-variables these are usually $\tau $-functions of integrable KP hierarchy
and usually satisfy ``generalized string equation"  $L^{\{V\}}_{-1}\tau  = 0$.
These facts make GKM an appealing object to study in the context of
non-perturbative and unified string theory.

This paper describes some new enlightening results about GKM. The question to
be addressed is what is the meaning of GKM from the point of view of the
Toda hierarchy. The point is that the KP hierarchy can be always embedded into
the Toda lattice hierarchy, but the latter one has additional variables:
``negative times" and ``zero-time", which are frozen in KP case. The natural
thing to ask is whether GKM can be further generalized to include the
dependence on these additional parameters so that non-perturbative partition
function becomes $\tau $-function of the Toda lattice hierarchy.

Remarkably it appears possible not only to introduce the extra
``zero-time" variable into GKM, but this is also a natural way to describe
{\it discrete} matrix models, thus giving a chance to unify the entire theory
of matrix models under a common roof of GKM.

The natural form in which the non-perturbative partition
functions\footnote{We use the term ``non-perturbative partition function"
for the generating
functional of {\it all} the {\it exact } ($i.e.$ non-perturbative or
appropriately summed over all orders of perturbation theory) correlation
functions in string models. This object may be considered as a vacuum amplitude
in the theory with the action, to which all the possible vertex operators
(corresponding both to naive observables and to handle-gluing operators) are
added with arbitrary coefficients (which are nothing but the time-variables).
Such quantity, though a priori constructed for one particular string model, a
posteriori is naturally describing a family of models as large as the freedom,
allowed in the deformations of the action. For a sufficiently rich family of
deformations the quantity, describing the entire string theory ($i.e.$
unification of all the string models), may be derived starting from any
particular string model.}
arise, when deduced from the matrix models, is determinant of $N\times N$
matrix, probably with some extra simple (normalization) factors in front of it.
Integrable $\tau $-functions are also representable through determinant
formulae, but somewhat different expressions are adequately describing
different hierarchies. This representation of KP $\tau $-function arises
naturally in Miwa parameterization of time-variables $t_p$, $i.e.$ essentially
in terms of eigenvalues $\{m_i\}$ of the $N\times N$ matrix $M$, and looks like

\beq
\tau ^{\{V\}}_{KP}[t] = {\det \ \phi _i(m_j)\over \Delta (m)},
\eeq
where  $\displaystyle{\{\phi _i(m) = m^{i-1}(1+{\cal O} ({1\over m}))\}}$ is a
Segal-Wilson
basis, describing some point in the Sato's Grassmannian (see all the details
and references in [1]). On the other hand, the natural determinant formula for
the Toda hierarchy is (see Appendix A, eq.(A.25))

\beq
\tau _n[t] = \det _{(ij)} H_{i+n,j+n}[t],
\eeq
where $H_{ij}$ satisfy the following equations:

\beq
\new
\begin{array}{c}
\partial H_{ij}/\partial t_p = H_{i,j-p}\hbox{ for ``positive times" } t_p,\\
\partial H_{ij}/\partial t_{-p} = H_{i-p,j}\hbox{ for ``negative times" }
t_{-p},
\end{array}
\eeq
and $n$ is integer-valued ``zero-time".

In what follows we need more statements from the Toda theory. The {\it forced}
Toda-lattice hierarchy arises if

\beq
\tau _{-n}[t] = \delta _{n,0}\hbox{ for all }\ n\geq 0.
\eeq
The reduction of general Toda lattice hierarchy to the Toda chain one occurs,
for example, when $H_{ij}$ depends only upon the {\it difference} $i-j$:
$H_{ij}
= {\cal H}_{i-j}$\footnote{The relations of this type, having a sense under
the determinant, should
be understood up to lower or upper triangle transformations admissible in the
determinant.}.
(More details and references concerning Toda hierarchies can be found in [4],
see also (2.20) below).

The partition function of GKM has been already proved in [1] to be a KP
$\tau $-function and can be explicitly represented in the form of (1.2). Our
first goal below will be to bring it to the form of (1.3). Usually KP hierarchy
can be embedded into Toda lattice hierarchy at fixed values of zero- and
negative-times\footnote{We systematically use the notation  $[\ldots]$ to
denote
the
dependence of a {\it family of arguments} like eigenvalues of $M$ or all the
time-variables (negative or positive). However, when the same objects are
considered as {\it parameters} we put them into braces: $\{\ldots\}$. For
example, negative times $t_{-p}$ are {\it arguments} of Toda lattice
$\tau$-function, but they are {\it parameters} of KP $\tau$-function (they
define the point in Grassmannian, but are not the time variables which are
conventionally considered as {\it arguments} of $\tau_{KP}$). The main purpose
of the study matrix models is certainly to get rid of this delicate difference,
$i.e.$ to treat all the {\it parameters} (which label different string models)
as {\it arguments}.}:

\beq
\tau _n[t_{-p};t_p] = \tau ^{\{n;t_{-p}\}}_{KP}[t_p]\hbox{ .}
\eeq
In other words, the point of Grassmannian associated with the KP
$\tau $-function at the $r.h.s.$ depends on  $t_{-p}$. This is also clear from
the free-field representation of $\tau $-functions (see Appendix A): a
Toda-lattice $\tau $-function,

\beq
\tau ^{\{g\}}_n[t_{-p};t_p] \sim  <n|e^{-\sum t_pJ_p}
g\ e^{-\sum t_{-p}J_{-p}}|n>
\eeq
can be considered as a KP $\tau $-function,

\beq
\tau ^{\{g;n;t_{-p}\}}_{KP}[t_p] \sim  <n|e^{-\sum t_pJ_p} g_{KP}|n>
\eeq
with $t_{-p}$-dependent  $g_{KP} = g\ e^{-\sum t_{-p}J_{-p}}$. In GKM the point
$g$ is specified by potential $V(X)$ (this relation is conventionally encoded
in the form of the ``string equation"). We shall see that $n$-
and $t_{-p}$-dependencies, as defined by eqs.(1.6) and (1.8), can be imitated
by additional logarithmic and negative-power terms in the action of GKM. This
is a sort of deformation of GKM in the sense that the original string equation
[1] still remains to be valid. A natural question is what are the restrictions
on GKM, which imply the occurrence of the {\it forced} Toda hierarchy.
Essentially these requirements are for all the functions $\phi _i(m)$ to be
{\it polynomial} in  $m$ (up to trivial factor, see, for example, eq.(3.24)),
and this is the case if $V(X) = X^2/2$. The GKM as
defined in [1] $(i.e$. with $n=t_{-p}=0)$ is trivial for such potential, but as
we shall see it becomes more interesting at $n\neq 0$ and
$t_{-p}\neq 0$\footnote{The particular model (with $t_{-p}\equiv 0$
but $n\neq 0)$ has been
proposed and discussed to some extent in a very recent paper [5], which
stimulated us to complete this research. Despite our disagreement with the main
claim of that paper (that GKM with $V(X)\sim X^2$ and $n>0$ describes $c=1$
string model) we shall refer in appropriate place to the proper Ward identity
(in fact, the system of {\it discrete} Virasoro constraints [6]), which is
derived there.}.

The second big problem addressed in this paper is the representation of
{\it discrete} matrix models in the form of GKM. One should suggest that such
representation exists if we indeed want GKM to be a kind of a {\it universal}
matrix model (encoding all the information about the KP hierarchy and
Grassmannian, which is relevant for string theory). This problem is also
related to the Toda theory, since partition functions of discrete matrix models
are known to be $\tau $-functions of {\it forced} Toda-lattice hierarchy (or
even its Toda-chain reduction in the one-matrix case) [6]. This problem is, in
fact, a kind of inverse of the previous one: now we need to convert the models
with characteristic Toda-type representation (1.3) into KP-related form (1.2),
which is peculiar for GKM. As usually, such conversion is provided by Miwa
transformation (1.1), moreover the resulting  $\{\phi _i(m)\}$ in (1.2) are
also orthogonal polynomials (up to trivial factor). As a particular result, we
explicitly prove the equivalence of {\it discrete} Hermitean one-matrix model
and
GKM with $V(X)\sim X^2$ and the zero-time  $n$  is identified with the size of
the matrix in the discrete model.

The fact that discrete matrix models can be naturally embedded into GKM, allows
one to formulate and study the ``double"-scaling continuum limits as internal
problem of GKM, where it becomes just a question about asymptotic formulas for
families of integrals. We present some naive results about such interpretation
of continuum limits, but its explicit relation to conventional Kazakov's
procedure (as described in full details in [7]) remains still a bit obscure.

\section{Partition function of GKM as a Toda-lattice $\tau $-function}
\setcounter{equation}{0}
\subsection{Interrelation between KP and Toda-lattice $\tau $-functions}

The purpose of this subsection is to describe without any special reference to
GKM the explicit relation between KP-like (in Miwa variables),

\beq
\tau _{KP}[t_p] = {\det _{ij}\phi _i(m_j)\over \Delta (m)},
\eeq
and Toda-like,

\beq
\tau _n[t_{-p},t_p] = \det _{ij} H_{i+n,j+n}[t_{-p},t_p],
\eeq
representations of $\tau $-functions, where

\beq
\Delta (m) =\prod _{i>j}(m_i-m_j),
\eeq

\beq
\phi _i(m) = m^{i-1}(1+{\cal O} ({1\over m})),
\eeq

\beq
t_p = {1\over p} \sum  _i m^{-p}_i\hbox{, }   p>0,
\eeq

\beq
\partial H_{ij}/\partial t_p = H_{i,j-p}\hbox{, }   j>p>0,
\eeq
and

\beq
\partial H_{ij}/\partial t_{-p} = H_{i-p,j}\hbox{, }   i>p>0.
\eeq
(the origin of these formulas, concerning with Toda hierarchy, is explained in
Appendix A).

Relation between (2.1) and (2.2) is formulated in terms of the Schur
polynomials, which are defined by:

\beq
{\cal P}[z|t_p] \equiv  \exp \{\sum _{p>0}t_pz^p\} = \sum    z^kP_k[t],
\eeq
$e.g.$  $P_{-n} = 0$  for any $n>0$;  $P_0[t] = 1$;  $P_1[t] = t_1$;  $P_2[t] =
t_2 + {1\over 2}t^2_1$;  $P_3[t] = t_3 + t_2t_1 + {1\over 6}t^3_1$ etc. The
crucial property of Schur polynomials is:

\beq
\partial P_k/\partial t_p = P_{k-p}
\eeq
(this is just because  $\partial {\cal P}/\partial t_p = z^p{\cal P})$. This
feature allows one to express all the dependence on time-variables of
$H_{ij}[t]$, which satisfies eqs.(2.6) and (2.7), through Schur polynomials
(see (A.26) in Appendix)\footnote{We manifestly write down the limits of this
sum only for convenience as
they are given automatically by properties of Schur polynomials.}:

\beq
H_{ij}[t_{-p},t_p] =
\sum _{^{k\leq i}_{l\geq -j}}P_{i-k}[t_{-p}]T_{kl}P_{l+j}[t_p],
\eeq
where  $T_{kl} = H_{kl}[0,0]$  is already a $t$-{\it independent} matrix. Note
that $H_{ij}$ is defined by eq.(2.10) for all (positive or negative) integer
values of $i,j$. Somewhat different, the Grassmannian point, $\{\phi _i(m)\}$,
in KP-formula (2.1) is {\it a priori} defined with $i\geq 0.$

Therefore, we begin our consideration from the case, when all  $n = t_{-p} =
0$, then look what happens if $n>0$, discuss the continuation to negative
values of $n$ and introduction of $t_{-p}$-variables. At the end of the
subsection we discuss the conditions for {\it forced} and/or Toda-{\it chain}
reductions to occur.

Given the system of basic vectors  $\phi _i(m)$  for $i>0$, we put by
definition

\beq
H_{ij}[t_{-p}=0,t_p] = \oint_{z \hookrightarrow  0}
\phi_i(z)z^{-j}{\cal P}[z|t_p]dz\hbox{, }      i>0\hbox{.}
\eeq
The integration contour is around zero and it can be deformed to encircle
infinity and the singularities of ${\cal P}[z]$, if any. If we just substitute
the definition (2.8) of ${\cal P}[z]$ into (2.11), we get (2.10) with
$P_{k-i}[t_{-p}=0] = \delta _{ki}$ and

\beq
T_{kl} = \oint_{z \hookrightarrow  0}
\phi_k(z)z^ldz.
\eeq
In order to prove the identity between (2.1) and (2.2) under the condition
(2.5), note that (2.5) implies that

$$
{\cal P}[z|t_p] = {\det \ M\over \det (M-Iz)} = \prod  _i {m_i\over (m_i-z)} =
\left[ \prod  _i m_i\right]  \sum  _k
{(-)^k\over (z-m_k)}{\Delta _k(m)\over \Delta (m)},
$$
where $\Delta _k(m) \equiv \prod _{\atop {i>j;}{i,j\neq k}}(m_i-m_j)$.
Note that
eigenvalues  $m_k = \infty $  do not contribute to ${\cal P} [z|t_p]$. If there
are exactly $N$ finite eigenvalues  $m_k\neq  \infty $, then the point
$z=\infty $ does not contribute to the integral (2.11), at least, for
$i\leq N$. It picks up contributions only from the poles of ${\cal P}[z|t_p]$
at the points $m_k$:

$$
H_{ij}[t_{-p}=0,t_p] = \oint_{z \hookrightarrow 0}
\phi_i(z)z^{-j}{\cal P}[z|t_p]dz = {
\prod_i m_i\over \Delta (m)} \sum  _k (-)^k \phi _i(m_k){\Delta _k(m)\over
m^j_k}.
$$
The sum at the $r.h.s.$ has a form of a matrix product and we conclude that

$$
\det \ H_{ij} = \det \ \phi _i(m_k)\cdot \prod  _k \left[ {
\prod _i m_i\over \Delta (m)} (-)^k \Delta _k(m)\right] \cdot \det
{1\over m^j_k}.
$$
The last determinant at the $r.h.s.$ is equal to  $\Delta (1/m) =
(-)^{N(N-1)/2} \Delta (m)\cdot \left[ \prod  _k m^N_k\right] ^{-1}$. Note also
that  $\displaystyle{\prod  _k \left[ {\Delta _k(m)\over \Delta (m)}\right]  =
\Delta (m)^{-2}}$, and gathering all this together, we see that the $r.h.s$. of
(2.12) is indeed equal to

\beq
\det \ H_{ij} = {\det \ \phi _i(m_j)\over \Delta (m)},
\eeq
as required.

Proceed now to introducing of zero- and negative-time variables. The zero-time
$n$  arises just as the simultaneous shifts of indices $i$ and $j$ of $H_{ij}:
H_{ij} \rightarrow  H_{i+n,j+n}$ , see (2.2). We can use eq.(2.11) to
write:

\beq
H_{i+n,j+n}[0,t_p] = \oint_{z \hookrightarrow 0}
\phi^{\{n\}}_i(z)z^{-j}{\cal P}[z]dz
\eeq
with

\beq
\phi ^{\{n\}}_i(z) = z^{-n}\phi _{i+n}(z)\footnote{Let us note that the
relations  $H^{(n)}_{ij} = H_{i+n,j+n}$ and
$\phi ^{(n)}_i = z^{-n}\phi _{i+n}$ are correct only for the special choice of
$H$ and $\phi $ (remind that they are defined up to triangle transformation).
Moreover, the choice of $H$ fixes $\phi $ unambigously due to eq.(2.11):
$H_{i+n,j+n} = \langle \phi ^{(n)}_iz^{-j}\rangle  =
\langle \phi ^{(n-k)}_{i+k}z^{-k}z^{-j}\rangle $  for any $j$, $i.e.$ {\it all}
moments of functions  $\phi ^{(n)}_i$ and  $\phi ^{(n-k)}_{i+k}z^{-k}$ are
coincide. This property, being evidently correct for GKM $\phi ^{(n)}_i$,
singles out the latter.}.
\eeq
This exhausts the problem of restoring the $n$-dependence for positive integer
values of $n$. As to $n<0$, eq.(2.15) can be used only if original set
$\{\phi _i(z)\}$  is enlarged to include $\displaystyle{\phi _i(z) =
z^{i-1}(1+{\cal O} ({1 \over z}))}$  with negative $i$. Such extension is not
limited by any additional
constraints and is not unique: this is exactly all the new information,
introduced when proceeding from KP to Toda-lattice hierarchy\footnote{In the
fermionic language of Appendix A this means that the form of KP
$\tau $-function allows one to restore the element $g$ of $GL(\infty )$, which
defines the point of Grassmannian, only up to a part, which cancels the vacuum
$|n\rangle$  ($n=0$ for conventional KP). Therefore, the choice of the negative
$n$ vacuum gives some additional information about $g$ in compare with $n=0$
(the same information can be, certainly, obtained from the form of the negative
time dependence). Thus, generally speaking, there are plenty ways to continue
$g$ to ``negative part". But it is not the case if one respects particular
reduction. For example, the condition of Toda chain reduction unambiguously
fixes the continuation to negative $n$ [4].}.
If this extension is chosen,  $H_{ij}[t_{-p}=0,t_p]$  and the entire matrix
$T_{kl}$ with {\it any} integer (positive or negative)  $i,j,k,l$  are defined.
{}From the point of view of Grassmannian the integer-valued zero-time  $n$
labels connected components of Grassmannian, consisting of the vector sets
$\{\phi _{i+n}(z)$, $i\geq 0\}.$

As for negative-times, as soon as $T_{kl}$ is defined, they are introduced with
the help of (2.10) and

\beq
\new
\begin{array}{c}
H_{i+n,j+n}[t_{-p},t_p] \equiv \sum _{k\leq i}P_{i-k}[t_{-p}]H_{k+n,j+n}[0,t_p]
=\\
= \oint_{z \hookrightarrow 0}
\phi^{\{t_{-p},n\}}_i(z)z^{-j}{\cal P}[{1\over z}|t_{-p}]{\cal P}[z|t_p]dz,
\end{array}
\eeq
with

\beq
\new
\begin{array}{c}
\phi ^{\{t_{-p},n\}}_i(z) \equiv
\left\lbrace {\cal P}[{1\over z}|t_{-p}]\right\rbrace ^{-1} \sum
P_{i-k}[t_{-p}]\phi ^{[n]}_k = \\
= z^{-n}\exp \left\lbrace -\sum _{p\geq 0}t_{-p}z^{-p}\right\rbrace  \sum
P_k[t_{-p}]\phi _{i+n-k}(z).
\end{array}
\eeq
The role of the exponential prefactor in (2.17) is to guarantee the proper
asymptotic behaviour

\beq
\phi ^{\{t_{-p},n\}}_i(z) = z^{i-1}\{1+{\cal O} ({1\over z})\}.
\eeq
Note that the quantities defined by eqs.(2.16) and (2.17) depend crucially on
$\phi _i$ with $i < 0$ when ``zero-time"  $n$  is negative.

Eqs.(2.16) and (2.17) provide a complete description of the interrelation
between KP and Toda-lattice hierarchies. Given a KP $\tau $-function in the
form of (2.1) ($i.e.$ in Miwa coordinates (2.5)), it can be interpreted as the
Toda-lattice $\tau $-function at the vanishing values of  $n$  and
$\{t_{-p}\}$. When zero-time  $n$  is introduced, it corresponds to discrete
shifts along disconnected components of Grassmannian, while evolution along
negative-times $t_{-p}$ is a smooth movement of a point in Grassmannian,
explicitly described by (2.17). (In other words, at any given values of  $n$
and $\{t_{-p}\}$ the Toda lattice $\tau $-function (2.2) as a function of
positive-times $\{t_p\}$ can be considered as KP $\tau $-function, but the
associated component and particular point of Grassmannian are different for
different  $n$  and $\{t_{-p}\}.)$

\bigskip
Further simplification of (2.17) can be achieved only for some particular
choices of the basis $\{\phi _i(z)$, $i\in {\bf Z}\}$. A drastic simplification
arises for basises, relevant for matrix models ($i.e.$ consistent with string
equations). Then  $\phi _i(z)$ are essentially of the form  $\left< x^{i-1}
\right> _z$
(with certain linear averaging operation $<\ldots>_z)$ (like contour integral
representation)\footnote{This interpretation of string
equation seems to be a very promissing view
on its meaning. Details are, however, beyond the scope of this paper.}
and introducing of  $n$  and $\{t_{-p}\}$ may be naturally described as a
change of the ``measure" from  $<\ldots>_z$ to

\beq
\left< \ldots
\left[ {x\over z}\right] ^n\exp \left\lbrace \sum _{p\geq 0}t_{-p}(x^{-p}-z^{-p
})\right\rbrace  \right> _z.
\eeq
Moreover, this formula allows one to consider  $n$  as continuous rather than
discrete parameter. This makes possible an ``analytic continuation" in  $n$
and, thus, implies a ``natural" definition of $\phi _i$ with negative  $i$.
Before we proceed to a more detailed discussion of this situation in the next
section, let us comment briefly on two important reductions of the Toda-lattice
hierarchy.

The first reduction which is of importance as it is just the case in matrix
models is already mentioned forced hierarchy. It was firstly introduced in
[8,4] for the Toda chain but can be easily extended to Toda lattice case. The
most manifest way to give this reduction is to constrain the element $g$ to
give some subspace in Grassmannian. This is done in Appendix $B$, but here we
would like to say some words in our previous framework.

That is, instead of half-infinite determinant in (2.2) let us consider finite
determinant of the size $n\times n$ for $\tau _n:$

\beq
\tau _n = \det_{n\times n}
H_{i,j}.
\eeq
It just gives a $\tau $-function of forced hierarchy. Moreover, due to
identities (2.6)--(2.7) it can be rewritten as\footnote{Let us
point out that this kind of solutions to Toda lattice hierarchy
was firstly invented by Leznov and Saveliev [9]. Let us also note that
solutions of Wronskian type [10] rather like forced ones turn out to be
absolutely different in their properties.}

\beq
\tau _n = \det_{n\times n}
\partial ^{i-1}\bar \partial ^{j-1}H
\eeq
with  $\partial \equiv \partial /\partial t_1$,
$\bar \partial \equiv \partial /\partial t_{-1}$, $H\equiv H_{1,1}$ and $H$ is
constrained to satisfy  $\partial H/\partial t_p  = \partial ^p H$,
$\partial H/\partial t_{-p}  = \bar \partial ^p H$. This $\tau $-function can
be produced in formalism of orthogonal polynomials implying polynomial
Baker-Akhiezer function as we shall see in the sect.3.1. It explains why matrix
models correspond to forced hierarchies.

For all these it is crucially to work in the sector with positive $n$. The
problem of continuation of forced hierarchies to negative zero-time we have
already discussed previously [4].

Another important reduction from Toda lattice is Toda chain
(see, for example, [10]). It can be easily
written both in terms of element $g$ ($[g,J_k +J_{-k}]=0$ -- see (A.29)
and comments there) and in determinant form. Latter one merely implies the
symmetry property:

\beq
[H\hbox{ , } \Lambda +\Lambda ^{-1}]=0,
\eeq
where $\Lambda $ is shift matrix  $\Lambda _{ij}\equiv \delta _{i,j-1}$. This
condition leads to $\tau $-function of Toda chain hierarchy (proper rescaled by
exponential of bilinear form of times) which depends only on the sum of
positive and negative times  $t_p+t_{-p}$\footnote{Let us emphasize that
usual notations correspond to the reduction to $tau$-function independent of
the sum of times. But due to non-standard sign of the first exponential in
(1.7)
which simplifies a lot of formulas in this paper, the reduction (2.25)
corresponds to $\tau$-function which depends only on the sum of times.},
but not on their difference (one can
consider this as defining property of Toda chain hierarchy). Let us remark that
one possible solution to constraint (2.22) is matrix
$H_{i,j}={\cal H}_{i-j}$\footnote{It
should not be certainly independent of the difference of times  $t_p-t_{-p}$,
but one can through out negative times as the final answer for complete objects
like $\tau $-function should be really independent of this difference.}.
We can combine both reductions to
reproduce forced Toda chain hierarchy. In this case one can easily transform
$H_{i-j}$
to matrix $\tilde {\cal H}_{i+j}$ by permutations of columns what does not
effect to the determinant. This matrix just corresponds to one-matrix model
case [4,6] (see also sect.3.1).
Thus, we consider again the
determinant of size $n\times n$, which now can be represent in the form:

\beq
\tau _n = \det_{n\times n}
\partial ^{i+j}H
\eeq
where $\partial \equiv \partial /\partial t_1$,
$\partial H/\partial t_p=\partial ^pH$  and we canceled negative times (see
footnote 12).
Like the Toda lattice case, the forced Toda chain is unambiguously
continuable to negative values of zero-time leading to (1.5)\footnote{This
continuation is unambigous only with taking into account the
negative times. With cancelled negative times, $i.e.$ in KP case, one can
obtain plenty different continuations like, for example, CKP $\tau $-function
which satisfies  $\tau _n[t_k] = \tau _{-n}[(-)^kt_k]$ [4].}.

\subsection{GKM in the context of Toda lattice hierarchy}

The purpose of this subsection is to introduce zero- and negative-time
variables into GKM in such a way that its partition function becomes a
$\tau $-function of the Toda lattice hierarchy. Let us remind the definition of
GKM at  $n = t_{-p} = 0$. It was introduced in [1] by the following
matrix integral:

\beq
Z_{\{V\}}[M] = {\int e^{-Tr\{V(X+M)-V(M)-V'(M)X\}}dX \over
\int e^{Tr\{-V_2(X,M)\}}dX}
\eeq
with $N\times N$ Hermitean matrices $X,M$ and  $V_2(X,M) \equiv
\lim_{\epsilon \rightarrow 0} \epsilon ^{-2}[V(M+\epsilon X)-V(M)-\epsilon
V'(M)X]$. As explained in
details in [1], the $r.h.s.$ of (2.24) may be represented in the form of (2.1)
and thus $Z_{\{V\}}[M]$ is the KP $\tau $-function, which has no explicit
dependence on $N$ ($N$ is just the number of {\it finite} eigenvalues of the
matrix $M$, when others can be considered as infinitely large). The relevant
set of functions $\{\phi _i(m)\}$ -- the point in Grassmannian -- is given by
the following integral formula:

\beq
\new
\begin{array}{c}
\phi ^{\{V\}}_i(m) = e^{V(m)-mV'(m)}\sqrt{V''(m)}\int
dx\ x^{i-1}e^{-V(x)+xV'(m)} \equiv \\
\equiv  s(m)\int   dx\ x^{i-1}e^{-V(x)+xV'(m)} \equiv  \left< x^{i-1}
\right> _m
\end{array}
\eeq
(the integral in (2.25) is a contour integral, $i.e.$ $x$ not a matrix).
Dependence of  $n$  and $t_{-p}$ is now introduced by the rule
(2.19)\footnote{Let us point out that the exponential of negative powers in
normalization does not essentially effect to the KP $\tau$-function as it
reduces to trivial exponential of bilinear form of times in front of
$\tau$-function and corresponds to the freedom in its definition [1].
Indeed, $\tau \sim \det \left\{ \exp
[\sum _ka_kz^{-k}_j]\phi _i(z_j)\right\}\sim \prod _l \exp [\sum a_kz^{-k}_l]
\det \phi _i(z_j) \sim \exp [\sum ka_kt_k] \det \phi _i(m_j)$. This factor,
certainly, effect to the string equation,
however, results into trivial bilinear form of times (see subsection 2.3,
eq.(2.37)).}:

\beq
\new
\begin{array}{c}
\phi _i ^{\{V,n,t_{-p}\}}(m) \equiv  \left<
x^{i-1}\left[ {x\over m}\right] ^n\exp \left\lbrace
\sum _{p\geq 0}t_{-p}(x^{-p
}-m^{-p})\right\rbrace  \right> _m = \\
= {\sqrt{V''(m)} e^{V(m)-mV'(m)}\over m^n}\int
dx\ x^{n+i-1}e^{-V(x)+xV'(m)}\exp \left\lbrace \sum _{p\geq
0}t_{-p}(x^{-p}-m^{%
-p})\right\rbrace  = \\
= e^{\hat V(m)-mV'(m)}\sqrt{V''(m)}\int   dx\ x^{i-1}e^{-\hat V(x)+xV'(m)},
\end{array}
\eeq
where

\beq
\hat V(X) \equiv  V(X) - n\log X - \sum _{p\geq 0}t_{-p}X^{-p} =
\sum _{p=-\infty }s_p {X^p\over p}\hbox{,}
\eeq
with

\beq
\new
\begin{array}{c}
t_p = {1\over p}[TrM^{-p} + (p-1)s_p],\\
t_0 = n = -s_0, \\
t_{-p} = -s_{-p}/p.
\end{array}
\eeq
As to original potential $V(m)$ it can be identified with $\hat V_+$. We repeat
that from the point of view of GKM  $t_0 = -s_0$ does not need to be integer
(though it is desirable to avoid a too complicated analytical structure of
$\hat Z[M]$ -- see (2.29) below).
{}From (2.26) we immediately conclude (just repeating the arguments
of [1] in the opposite direction) that the partition function of GKM, involving
zero- and negative-times (and automatically being a Toda lattice
$\tau $-function), is just\footnote{All the
integrals are certainly defined by analytic continuation and are
not quite unambigous because of Stokes phenomenon etc.}

\beq
\hat Z_{\{\hat V\}}[M] = e^{Tr\hat V(M)-TrM\hat V'_+(M)}
{\int dX\  e^{-Tr\hat V(X)+Tr\hat V'_+(M)X}\over
\int dX\ e^{-Tr\hat V_{+,2}(X,M)}}\footnote{Note also that $\hat Z_{\{
\hat V\}}[M] \neq  Z_{\{\hat V\}}[M]$ because of occurrence of
$\hat V_+$ at some places. One more model can
be defined by the rule ${\cal F}_{\{\hat V\}}[\hat V'_+(M)] = {\cal F}_{\{
\hat V\}}[\hat V'(\hat M)]$ (where ${\cal F}$'s denote only integrals in the
numerators of (2.24) and (2.29)). Then $\hat M$ is defined by the relation
$\hat V'_+(M) (= V'(M)) = \hat V'(\hat M)$, and $Z_{\{\hat V\}}
[\hat M]$ is a KP $\tau $-function of somewhat different time variables
$\displaystyle{\hat t_p = {1\over p}[Tr\hat M^{-p} + (p-1)\hat s_p]}$,
where $\hat s_p$ are
defined to be the coefficients of $\hat V$ expansion in powers of $\hat M$. In
the simplest case of  $\displaystyle{\hat V(X) = {1\over 2}X^2}-n \log X$  the
definition of $\hat M$ is  $M = \hat M - n\hat M^{-1}$. This is a rather
familiar relation in Toda chain theory connecting its spectral parameter with
that of KP hierarchy. We emphasize that while both $\hat Z_{\{\hat V\}}
[M]$ and $Z_{\{\hat V\}}[\hat M]$ are KP $\tau $-functions, only $\hat Z_{\{
\hat V\}}$ is also a Toda lattice $\tau $-function ($i.e.$ possesses a proper
dependence of $t_0$ and $t_{-p}$).}.
\eeq
Occurrence of GKM in the somewhat unexpected form of eq.(2.29) (involving
projection  $\hat V \rightarrow  \hat V_+)$ makes the whole subject even more
intriguing than it was before. Our next purpose is to describe the associated
modification of the second crucial ingredient of GKM: the string equation (the
first ingredient is integrability).

\subsection{Generalized string equation}

Two ways to derive string equation and $W$- (in particular, Virasoro)
constraints were suggested in [1]. One immediately leads to the string equation
$L^{\{V\}}_{-1}Z = 0$  and, as a consequence of it together with an additional
information about $Z$ (that it is a $\tau $-function of a specifically reduced
KP hierarchy), one reproduce the set of all other $W$-constraints appropriate
for the given reduction [11]. Another way [12] starts with exhaustive matrix
Ward identity [13]\footnote{See also [14] and [15].}
which is transformed then into the set of appropriate $W$-constraints, without
addressing to properties of $Z$ at all. Below we concentrate on the first
approach which is much simpler in general situation.

A $\tau $-function is parameterized by a point of Grassmannian.
The role of string
equation is to specify this dependence for the case of $\tau $-function given
by the GKM partition function. Unlike the Toda case, if we look at $Z_{\{\hat V
\}}$ as at KP $\tau $-function, the point of Grassmannian depends on entire
potential  $\hat V(X)$, $i.e.$ not only on the polynomial piece  $V(X) =
\hat V_+(X)$ but as well as on its negative and zero times (being included in
definition of the point of Grassmannian in KP framework).

\bigskip
In this subsection we shall shortly repeat the derivation of string
equation\footnote{In
this section we really deal with $L_{-1}$-constraint, the derivative
of which over the first time variable is usually named string equation.
Nevertheless, for the brevity, we call this Virasoro constarint as string
equation too.}
addressing reader for more details to [1]. We shall demonstrate that this
derivation crucially uses {\it only} the following information:

1) given asymptotics of  $\phi _i(\mu ) \stackreb{\mu \rightarrow \infty}{\sim}
 \mu ^{i-1}$;

2) the manifest form of normalization factor depending on $M$ in GKM integral;

3) the manifest form of linear term in $X$ in exponential in integrand (2.25)
(it is necessary in order to get correct expression (2.1) for $Z$ implying $Z$
to be $\tau $-function).

The main idea is to consider the following derivative of the GKM partition
function  $\displaystyle{Z = {det\phi _i(m_j)\over \Delta (m)} \equiv
{\det \ s(m_j)\tilde \phi _i(m_j)\over \Delta (m)}}$ (see (2.24)--(2.25)):

\beq
Tr\left\lbrace {1\over V''(M)} {\partial \over \partial M}
\log \ Z\right\rbrace
\eeq
and rewrite it as

\beq
\sum _{p>0}Tr {1\over V''(M)} {\partial t_p\over \partial M}
{\partial log\ Z\over \partial t_p} = - \sum _{p>0}Tr {1\over V''(M)}
{1\over M^{p+1}} {\partial log\ Z\over \partial t_p}\hbox{ .}
\eeq
On the other hand, the derivative (2.30) is equal to two pieces: the first one
originates from the derivative of factors $s(m)$ and $\Delta (m)^{-1}$ (here we
use the information of point 2) and is equal to:

\beq
{1\over 2} \sum _{i,j}{1\over V''(m_i)V''(m_j)} {V''(m_i) - V''(m_j)\over m_i -
m_j}\hbox{ .}
\eeq
The remaining second piece can be transformed to derivative over $t_1$
essentially using the correct asymptotics of $\phi _i(m)$ (point 1):

\beq
Tr\left\lbrace {1\over V''(M)} {\partial \over \partial M} \log \ \det
\tilde \phi _i(m_j)\right\rbrace  = {\partial \over \partial t_1}
\log \ Z\hbox{ .}
\eeq
Now let us modify GKM partition function by introducing of negative- and
zero-time variables, in accordance with (2.29). Then the only change (one
should certainly to replace all $V$ by $\hat V_+)$ of string equation
originates from the additional piece in (2.32):

\beq
Tr {1\over \hat V''_+(M)} {\partial \hat V_-(M)\over \partial M} \log \ Z
= - t_0 Tr {1\over \hat V''_+(M)M}
+\sum _{p>0} pt_{-p} Tr {1\over \hat V''_+(M)M^{p+1}} .
\eeq
Thus, we finally obtain the string equation in the form:

\beq
\new
\begin{array}{c}
L_{-1}\tau  = 0,\\
L^{\{\hat V_+\}}_{-1} = \sum _{p>0}{\cal T}_{p+1}^{\{\hat V^+\}}
{\partial \over \partial t_p}
+ {\partial \over \partial t_1} - {\cal T}_1n^{\{\hat V^+\}} +
\sum _{p>0} p{\cal T}_{p+1}^{\{\hat V^+\}}t_{-p}
+\\
+ {1\over 2} \sum _{i,j}{1\over
\hat V''_+(m_i)\hat V''_+(m_j)}
{\hat V''_+(m_i) - \hat V''_+(m_j)\over m_i - m_j}
\end{array}
\eeq
with

\beq
{\cal T}^{\{\hat V_+\}}_p \equiv  Tr {1\over
\hat V''_+(M)M^p}.
\eeq
Let us note again (see footnote 14) that it is not necessary to
include negative times into normalization factor $\hat s(m)$. In this case, the
string equation is not at all modified by negative times. Moreover, we can
multiply GKM integrand by exponential of arbitrary series  $\sum ^k_{-\infty }
a_pX^p$ with the only restriction  $k <$ (the leading degree of $\hat V(X)$) to
preserve the property 1 of correct asymptotics of $\phi _i$. One can easily see
from our previous discussion that this does not disturb three essential
properties above and the form of the string equation is conserved. Thus, in KP
framework we have plenty deformations of starting point of Grassmannian not
changing the string equation. These modifications does certainly change the
reduction condition, therefore, it emphasize the importance of fixing the
reduction to define uniquely the  $\tau $-function of (reduced) KP hierarchy by
the string equation, in spirit of [11].

On the other hand, one can merely consider this as an extension of standard GKM
viewpoint. Nevertheless, usual GKM representation is singled out by the
possibility to do smooth transition to potentials of the other degrees
corresponding to the double scaling continuum limit of the other multi-matrix
models. In this sense, one can look at the string equation as giving some sort
of universality class.

Now let us return to the string equation (2.35). For particular potential
$\hat V_+(X) = X^{K+1}/(K+1)$ ,  ${\cal T}_{p+1} = (p+K)t_{p+K}$ and
$L_{-1}$-constraint has a form:

\beq
L^{\{K\}}_{-1} = \sum _{p>0} (p+K)t_{p+K} {\partial \over \partial t_p} + K
{\partial \over \partial t_1} + {1\over 2} \sum _{a+b=K}at_abt_b
\eeq
with $a,b$ being any (positive and negative) integers and $at_a$ should be
substituted by $n$ for $a=0$. For all $t_{-p}=0$ the term
$\partial /\partial t_1$ can be interpreted as resulting from the shift
$t_p\rightarrow t_p + \displaystyle{K\over K+1}
\delta _{p,K+1} \equiv  \hat t_p$. If also
$K=1,$

\beq
L^{\{1\}}_{-1} = \sum _{p>0} (p+1)\hat t_{p+1} {\partial \over \partial t_p} +
nt_1
\eeq
and this is similar to the string equation of discrete Hermitean one-matrix
model, $n$ being the size of the matrix. We shall return to this point a bit
later. Here only note that, while in [5,6,16] it was proposed to interpret $n$
as  $\partial /\partial t_0$, we see now it is more natural identify $n$ with
$t_0$ itself, in the sense that it is $n$ that plays the role of zero-time in
Toda hierarchies (this was mentioned but not emphasized in [6]).

Somewhat alternative approach to the derivation of string equation and
associated tower of $W$-like constraints could be to begin with the Ward
identity for the integral

\beq
F(\Lambda ) = \int   dX\ e^{-Tr\hat V(X) + Tr\Lambda X}\hbox{ ,}
\eeq
which results from the invariance under a shift  $X\rightarrow X+\epsilon$
($\epsilon $  being a matrix) of the integration variable:

\beq
\langle Tr\ \epsilon \{\hat V'(X) - \Lambda \}\rangle  = 0.
\eeq
Usually we can represent the positive powers of $X$ coming from  $\hat V'_+(X)$
in the integrand (before averaging) in (2.40) by action of derivative operator
$\hat V'_+(\partial /\partial \Lambda _{tr})$  to already averaged expression.
This is naively impossible for negative powers of $X$ in  $\hat V'_-(X)$.
However, be there {\it only finite number} of negative powers ($i.e.$ if
$\hat V'_-(X)$ is a {\it polynomial} in $X^{-1}$), one could take additional
$\Lambda $-derivatives of equation (2.40) in order to obtain more factors of
$X$ and eliminate its negative powers. This trick was applied in [5] in the
particular case of all $t_{-p}=0$ with only non-zero $n$. In this way they
obtained the identity of the form:

\beq
Tr \epsilon \left\lbrace {\partial \over \partial \Lambda _{tr}}
\hat V'_+\left[ {\partial \over \partial \Lambda _{tr}}\right]  +
{\partial \over \partial \Lambda _{tr}}\Lambda  + n\right\rbrace  F(\Lambda ) =
0
\eeq
which, if rewritten in terms of $Z$ and in the limit of $N\rightarrow \infty $
(see details im [12] and [5]) turns into a set of Virasoro constraints

$$
L_pZ = 0\hbox{, } p\geq -1
$$
with

\beq
\new
\begin{array}{c}
L_p = \sum _{k=1} k\hat t_k {\partial \over \partial t_{k+n}} +
\sum ^{n-1}_{k=1} {\partial ^2\over \partial t_k\partial t_{n-k}} + 2n
{\partial \over \partial t_p}\hbox{  for }\ p\geq 1,\\
L_0 = \sum _{k=1} k\hat t_k {\partial \over \partial t_k} + n^2\hbox{
and }\ \ L_{-1}\hbox{ as in (2.38),}
\end{array}
\eeq
which are identified with Virasoro constraints for Hermitean one-matrix model.
If some finite number of negative times are non-vanishing, one get instead a
set of $\tilde W$-constraints [17].

\section{Discrete models in the form of GKM}
\setcounter{equation}{0}

Since we devote this paper to discussion of Toda lattice hierarchies in the
context of matrix models, we can not avoid touching the main conclusion of
ref.[6] that all the {\it discrete} matrix models do correspond to particular
cases of Toda hierarchies. In the simplest case of Hermitean one-matrix model
one
gets a Toda chain, other multi-matrix models correspond to other reductions of
the Toda lattice hierarchy. Moreover, all discrete matrix models fall into the
class of {\it forced} hierarchies [4].

\subsection{Discrete models as forced Toda-lattice $\tau $-functions}

In this subsection we show how partition functions of various matrix models can
be rewritten in the KP-like form of eq.(2.1).

The first step in this direction is to reproduce them in the Toda like form of
(2.2). This statement was already discussed in [6] and [4], but we present it
more explicitly below. The second step is to notice that  $H_{ij}$ which arises
from discrete models are representable as averages, namely, finite dimensional
integrals, or ``matrices of moments":

$$
H_{ij}[t_{-p},t_p] = \langle x^iy^j{\cal P} [x|t_p]{\cal P} [y|t_{-p}]\rangle .
$$
The third step is to perform a Miwa transformation of times $t_p$ with $t_{-p}$
fixed, so that  $H_{ij}$'s become averages of polynomial functions of {\it X}.
Then  $\det \ H_{ij}$ may be transformed with the help of orthogonal
polynomials technique. These polynomials are, however, orthogonal with respect
to the simple measure, defined by potential  $\tilde V(H) \equiv  V(H) -
HV'(H)$, which depends only on the  $s_p$-variables and not on the times
$t_p$.

The main result of all these calculations is that partition functions arise in
the form (2.1), with  $\phi _i(z)$  being proportional to orthogonal
polynomials.

First, we shall briefly repeat what is known about discrete matrix models. In
the case of the Hermitean one-matrix model the $n\times n$ matrix integral

\beq
\new
\begin{array}{c}
Z^{(1)}_n = \{Vol_{U(n)}n!\}^{-1}\int   DH\ \exp \{ -\sum    t_{_k}SpH^{_k}\}
= \\
= (n!)^{-1}\int   \prod  _i dh_i \Delta ^2(h) \exp \{
-\sum _{i,k}t_{_k}h^{_k}_i\}
\end{array}
\eeq
is taken by orthogonal polynomials for the arbitrary potential  $W(h) = \sum
t_kh^k$

\beq
<P^{(1)}_i,P^{(1)}_j> = \int   P^{(1)}_i(h)P^{(1)}_j(h)e^{-W(h)}dh =
\delta _{ij}e^{\varphi _i(t)}
\eeq
and equals

\beq
Z^{(1)}_n = \prod ^{n-1}_{i=0}e^{\varphi _i(t)} = \tau ^{(1)}_n(t)
\eeq
which is the  $\tau $-function of the forced Toda chain hierarchy  (being
defined for negative  $n$  by (1.5)). In eq.(3.1) $Vol_{U(n)}$ denotes the
volume of unitary group $U(n)$ (see (3.41) for explicit expression).
On the other hand, it follows from (3.2)
and the definition of orthogonal polynomials

\beq
\new
\begin{array}{c}
P^{(1)}_i(h) = \sum _{j\leq i}a^{(1)}_{ij}h^j \\
a^{(1)}_{ii} = 1
\end{array}
\eeq
that

\beq
diag(e^{\varphi _i(t)}) = A^{(1)}C^{(1)}A^{(1)T} ,
\eeq
where  $A^{(1)} = \|a^{(1)}_{ij}\|$,  $A^T$ -- transponed matrix, and
$C^{(1)}$
is so called matrix of moments

\beq
C^{(1)}_{ij}= \int   h^{i+j}e^{-W(h)}dh.
\eeq
Thus [4],

\beq
\tau ^{(1)}_n(t) = \det [\hbox{diag}(e^{\varphi _i(t)})] =
\det \ A^{(1)}C^{(1)}A^{(1)T} = \det \ C^{(1)} \hbox{ . }
\eeq
More or less the same is true for all discrete matrix models, only the form of
{\it A} (which is unessential in (3.7)) and scalar product in the definition of
$C$  depend on the model. In the case of two Hermitean matrices and two
potentials  $W(h) = \sum    t_kh^k$ and  $\bar W(\bar h) = \sum
\bar t_k\bar h^k$ we have

\beq
\new
\begin{array}{c}
Z^{(2)} = \{Vol_{U(n)}n!\}^{-1}\int   DH\ D\bar H e^{-Sp[W(H) + \bar W(\bar H)
- H\bar H]} =\\
= (n!)^{-1}\int   \prod  _i dh_i d\bar h_i
\Delta (h)\Delta (\bar h) \exp \{-\sum  _i [W(h_i) + \bar W(\bar h_i) -
h_i\bar h_i]\} = \\
= \prod ^{n-1}_{i=0}e^{\varphi _i(t,\bar t)} = \tau ^{(2)}_n(t,\bar t)\hbox{ .
}
\end{array}
\eeq
where

\beq
e^{\varphi _i(t,\bar t)}\delta _{ij} = <P^{(2)}_i,\bar P^{(2)}_j> = \int
P^{(2)}_i(h)\bar P^{(2)}_j(\bar h)e^{h\bar h-W(h)-\bar W(\bar h)}dhd\bar h
\eeq
and

\beq
\tau ^{(2)}_n(t,\bar t) = \det [\hbox{diag}(e^{\varphi _i(t,\bar t)})] =
\det \ A^{(2)}C^{(2)}\bar A^{(2)T} = \det \ C^{(2)}
\eeq
for matrix of moments given by

\beq
C^{(2)}_{ij}= \int   h^i\bar h^je^{h\bar h-W(h)-
\bar W(\bar h)}dhd\bar h\hbox{ . }
\eeq
In generic  $K$-matrix situation we can also use the formulas (3.10) and (3.11)
if we take care of the integration measure. Namely,

\beq
\new
\begin{array}{c}
Z^{(K)}_n = \{Vol_{U(n)}n!\}^{-1}\int   DH\ D\bar H \prod ^{K-1}_{l=2}DH^{(l)}
\exp \{- Sp\{W(H) + \bar W(\bar H) + \sum  ^K_2 V^{(l)}(H^{(l)}) - \\
- \sum    c_lH^{(l)}H^{(l+1)} - HH^{(2)} - H^{(K-1)}\bar H\}\} =      \\
= (n!)^{-1}\int   \prod  _i dh_i d\bar h_i \prod ^{K-1}_{l=2}dh^{(l)}_i
\Delta (h)\Delta (\bar h) \exp \{-\sum  _i [W(h_i) + \bar W(\bar h_i) +
\sum ^{K-1}_2W^{(l)}(h^{(l)}_i) - \\
- \sum    c_lh^{(l)}_ih^{(l+1)}_i- h_ih^{(2)}_i - h^{(K-1)}_i\bar h_i]\} = \\
= \prod ^{n-1}_{i=0}e^{\varphi _i(t,\bar t;W^{(l)},c_l)} =
\tau ^{\{K;W^{(l)},c_l\}}_n[t,\bar t] \hbox{ , }
\end{array}
\eeq
where we distinguished  $H^{(1)}\equiv H$  and  $H^{(K)}\equiv \bar H$  among
other integration variables, because the role of Toda time variables is played
by  $t^{(1)}\equiv t$  and  $t^{(K)}\equiv \bar t$, while other times
$t^{(l)};$
$l = 2,\ldots,K-1$  (or potentials  $W^{(l)}$) and the set of $\{c_l\}$  are
considered as parameters. (The evolution along these frozen times can be
described by the {\it multi}-{\it component} Toda lattice hierarchy.)

The orthogonal polynomials now look like

\beq
\new
\begin{array}{c}
e^{\varphi _i(t,\bar t;\{W\},\{c\})}\delta _{ij} = <P^{(K)}_i,\bar P^{(K)}_j>
= \\
= \int   P^{(K)}_i(h)\bar P^{(K)}_j(\bar h)d
\mu ^{\{K;\{W\},\{c\}\}}[h,\bar h]\hbox{ , }
\end{array}
\eeq
where

\beq
\new
\begin{array}{c}
d\mu ^{(K)}(h,\bar h) = dhd\bar he^{-Sp\{W(h) + \bar W(\bar h)\}}\int
\prod ^{K-1}_2dh^{(l)}\exp \{- Sp\{\sum ^{K-1}_2W^{(l)}(h^{(l)}) - \\
- \sum    c_lh^{(l)}h^{(l+1)} - hh^{(2)} - h^{(K-1)}\bar h\}\}\hbox{ . }
\end{array}
\eeq
Again

\beq
\new
\begin{array}{c}
\tau ^{\{K;\{W\},\{c\}\}}_n[t,\bar t] =
\det [\hbox{diag}(e^{\varphi _i(t,\bar t;\{W\},\{c\})})] = \\
= \det \ A^{(K)}C^{(K)}\bar A^{(K)T} = \det \ C^{(K)}
\end{array}
\eeq
and the matrix of moments is

\beq
C^{(K)}_{ij}= \int   h^i\bar h^jd\mu ^{\{\{W\},\{c\}\}}[h,\bar h]\hbox{ . }
\eeq
In the limit  $K \rightarrow  \infty $  this should correspond to the
infinite-dimensional matrix model, whose first critical point has to describe
(at least, formally)  $c = 1$  conformal matter coupled to gravity [18].
Instead
of the infinite sum we introduce the ``quantum mechanical" integral and (3.12),
(3.14) can be rewritten as

\beq
\new
\begin{array}{c}
Z^{(\infty )}_n \sim  \int ^{H(\Xi )=\bar H}_{H(0)=H}\prod  _\xi  DH(\xi )
\exp \{- Sp\int _0^{\Xi }d\xi \left\{ W[H(\xi )] +
\left( {\partial H\over \partial \xi }\right) ^2\right\}\} = \\
= \int   \prod  _i d\mu ^{(\infty )}(h_i,\bar h_i) \Delta (h)\Delta (\bar h)
= \\
= \prod ^{n-1}_{i=0}e^{\varphi _i[t,\bar t;W(\xi )]} =
\tau ^{\{\infty ;W(\xi )\}}_n[t,\bar t]
\end{array}
\eeq
with

\beq
d\mu ^{(\infty )}(h,\bar h) = dhd\bar h \int ^{h(\Xi )=\bar h}_{h(0)=h}\prod
_\xi  Dh(\xi ) \exp \{- Sp\int _0^{\Xi} d\xi \left\{ W[h(\xi )] +
\left( {\partial h\over \partial \xi }\right) ^2\right\}\} \hbox{ . }
\eeq
So far we considered the partition functions of various discrete matrix models
as functions of Toda time-variables. Now we shall demonstrate that if one
passes from times to Miwa variables

\beq
t_p = {1\over p} TrM^{-p} = {1\over p}\sum ^N_{i=1}m^{-p}_i + {p-1\over p} s_p
\eeq
($p > 0$) and/or

\beq
t_{-p} \equiv  \bar t_p = {1\over p}Tr\Lambda ^{-p} = {1\over p}\sum ^N_{i=1}
\lambda ^{-p}_i + {p-1\over p} \bar s_p
\eeq
(note that  $N$ -- the size of matrices  $M$  and  $\Lambda $  has nothing to
do
with  $n$ --
the size of matrices  $H$, being integrated over in (3.1), (3.8) and
(3.12)) then the partition functions of discrete models  $\{\tau ^{(K)}_n\}$
acquire the form of KP $\tau $-function (2.1). Indeed,

\beq
\new
\begin{array}{c}
\tau ^{(1)}_n(t) = (n!)^{-1}\int   \prod  _i dh_i \Delta ^2(h) \exp
\{-\sum _{i,k}t_{_k}h^{_k}_i \} = \\
= (n!)^{-1}\int   \prod  _i dh_i \Delta ^2(h) e^{-\tilde V(h_i)}
\prod _{i,a}(1 -
{h_i\over m_a}) = \\
=(n!)^{-1}\prod  _a m^{-n}_a\int   \prod  _i dh_i e^{-\tilde V(h_i)} \Delta (h)
{\Delta (h,m)\over \Delta (m)} = \\
= (n!)^{-1}\prod  _a m^{-n}_a \Delta ^{-1}(m)\int   \prod  _i dh_i
e^{-\tilde V(h_i)} \times \\
\times  \det _{n\times n} \tilde P^{(1)}_{i-1}(h_j) \det _{(n+N)\times (n+N)}
\left[ \begin{array}{rcl}
\tilde P^{(l)}_{i-1}(h_j) & \vdots & \tilde P^{(l)}_{n+b-l} (h_j)\\
\ldots & \vdots & \ldots\\
\tilde P^{(l)}_{i-1}(m_a) & \vdots & \tilde P^{(l)}_{n+b-l} (m_a)
\end{array}
\right]\hbox{ , }
\end{array}
\eeq
where  $i,j = 1,\ldots,n$;  $a,b = 1,\ldots,N$;  $\displaystyle{\tilde V(h)
= \sum
{k-1\over k}s_kh^k}$ and  $\{\tilde P^{(1)}_i(h)\}$  are corresponding
orthogonal polynomials with respect to deformed measure  $e^{-\tilde V} dh:$

\beq
<\tilde P^{(1)}_i,\tilde P^{(1)}_j> = \int
\tilde P^{(1)}_i(h)\tilde P^{(1)}_j(h)e^{-\tilde V(h)}dh =
\delta _{ij}e^{\tilde \varphi _i(s)}\hbox{ , }
\eeq
so that

$$
\tilde P^{(1)}_i(h) = h^i + O(h^{i-1}).
$$
Computing determinants in (3.21) and using orthogonality condition (3.22) one
obtains

\beq
\new
\begin{array}{c}
\tau ^{(1)}_n[m|s] = \prod  _a m^{-n}_a \Delta ^{-1}(m) \det _{N\times N}
\tilde P^{(1)}_{n+a-1}(m_b) \prod  _i e^{\tilde \varphi _i(s)} = \\
= \left[ \prod  _i e^{\tilde \varphi _i(s)}\right]
{\det _{(ab)}\phi ^{(1,n)}_a(m_b)\over \Delta (m)} = \tau ^{(1)}_n[\infty |s]
\times {\det _{(ab)}\phi ^{(1,n)}_a(m_b)\over \Delta (m)}\hbox{ ,}
\end{array}
\eeq
$i.e.$ the $\tau $-function of the discrete Hermitean one-matrix model acquires
the form of eq.(2.1) with

\beq
\phi ^{(1,n)}_a(m) = m^{-n} \tilde P^{(1)}_{n+a-1}(m)\hbox{ . }
\eeq
Below we shall see that (3.24) is natural representation for all discrete
matrix models.

Let us remark that the expressions (3.21)-(3.24) does not depend on the
quantity {\it N}. It means that one can consider eq.(3.21) with $N=1$ and
reproduce well-known integral representation for orthogonal polynomials [19]
(the serious drawback of this representation is that its manifest form, and,
moreover, the number of integrations depend on the degree of polynomial):

$$
\tilde P^{(1)}_n(m) =  {\int \prod _idh_i \Delta ^2(h)e^{-\tilde V(h)}
\prod _i(m - h_i)\over
\int \prod _idh_i \Delta ^2(h) e^{-\tilde V(h)}}\hbox{ .}
$$

\bigskip
In the case of two matrices instead of (3.21) one gets

\beq
\new
\begin{array}{c}
\tau ^{(2)}_n(t,\bar t) = (n!)^{-1}\int   \prod  _i dh_i d\bar h_i
\Delta (h)\Delta (\bar h) \exp \{-\sum  _i [W(h_i) + \bar W(\bar h_i) -
h_i\bar h_i] \}= \\
= (n!)^{-1}\int   \prod  _i dh_i d\bar h_i \Delta (h)\Delta (\bar h) \exp
\{-\sum  _i [\tilde V(h_i)+\bar V(\bar h_i)-h_i\bar h_i]\}\prod _{i,a}(1 -
{h_i\over m_a}) = \\
=(n!)^{-1}\prod  _a m^{-n}_a\int   \prod  _i dh_i d\bar h_i \exp \{-\sum  _i
[\tilde V(h_i)+\bar V(\bar h_i)-h_i\bar h_i]\} \Delta (\bar h)
{\Delta (h,m)\over \Delta (m)} = \\
= (n!)^{-1}\prod  _a m^{-n}_a \Delta ^{-1}(m)\int   \prod  _i dh_i d\bar h_i
\exp \{-\sum  _i [\tilde V(h_i)+\bar V(\bar h_i)-h_i\bar h_i] \} \times \\
\times  \det _{n\times n} \tilde {\bar P}
^{(2)}_{i-1}(\bar h_j) \det _{(n+N)\times (n+N)} \left[
\begin{array}{rcl}
\tilde P^{(2)}_{i-1}(h_j) & \vdots & \tilde P^{(2)}_{n+b-l} (h_j)\\
\ldots & \vdots & \ldots\\
\tilde P^{(2)}_{i-1}(m_a) & \vdots & \tilde P^{(2)}_{n+b-l} (m_a)
\end{array}
\right]  = \\
= \left[ \prod  _i e^{\tilde \varphi _i(s,\bar t)}\right]
{\det _{(ab)}\phi ^{(2,n)}_a(m_b)\over \Delta (m)}\hbox{ , }
\end{array}
\eeq
where

\beq
\phi ^{(2,n)}_a(m) = m^{-n} \tilde P^{(2)}_{n+a-1}(m)\ ,
\eeq

\beq
<\tilde P^{(2)}_i,\tilde {\bar P}
^{(2)}_j> = \int   \tilde P^{(2)}_i(h) \tilde {\bar P}
^{(2)}_j(\bar h)e^{h\bar h-\tilde V(h)-\bar V(\bar h)}dhd\bar h =
e^{\tilde \varphi _i(s,\bar t)}\delta _{ij}\hbox{ . }
\eeq
Note that, using symmetry of the two-matrix model, one can obtain the formula
(3.25) using Miwa-transformed form of the other set of times (3.20), though its
proper interpretation is not yet completely clear. In this case:

\beq
\tau ^{(2)}_n(t,\bar t) = \left[ \prod  _i
e^{\tilde \varphi _i(s,\bar t)}\right]
{\det _{(ab)}\bar \phi ^{(2,n)}_a(\lambda _b)\over \Delta (\lambda )}
\eeq
with

\beq
\bar \phi ^{(2,n)}_a(\lambda ) = \lambda ^{-n} \tilde {\bar Q}
^{(2)}_{n+a-1}(\lambda )
\eeq
where

\beq
<\tilde Q^{(2)}_i,\tilde {\bar Q}
^{(2)}_j> = \int   \tilde Q^{(2)}_i(h)  \tilde {\bar Q}
^{(2)}_j(\bar h)e^{h\bar h-\tilde V(h)- \tilde {\bar V}
(\bar h)}dhd\bar h = e^{\tilde \varphi _i(t,\bar s)}\delta _{ij}
\eeq
for  $\displaystyle{\tilde {\bar V}
(\bar h) = \sum     {k-1\over k}\bar s_k\bar h^k}$.

Let us point out that formulas (3.25), (3.28) look especially nice in the
particular case of {\it asymmetric} two-matrix model [17,20] with one of the
potential being a finite polynomial of fixed degree, say
$\displaystyle{\bar W(\bar h) =
\sum ^M_{k=1}{k-1\over k}s_k\bar h^k}$. Then it is natural to take  $\tilde
V(h)
= W(h)$  and (3.25) turns to be

\beq
\tau ^{(2,W)}_n(m,s) = \left[ \prod  _i e^{\tilde \varphi _i(s,s)}\right]
{\det _{(ab)}\phi ^{(2,n,W)}_a(m_b)\over \Delta (m)}\hbox{ , }
\eeq
where

\beq
\phi ^{(2,n,W)}_a(m) = m^{-n} P^{(2,W)}_{n+a-1}(m)\ ,
\eeq

\beq
\new
\begin{array}{c}
<P^{(2,W)}_i,\bar P^{(2,W)}_j> = \\
= \int
P^{(2,W)}_i(h)\bar P^{(2,W)}_j(\bar h)e^{h\bar h-W(h)-W(\bar h)}dhd\bar h =
e^{\varphi _i(s,s)}\delta _{ij}.
\end{array}
\eeq
As we have seen above orthogonal polynomials technique without serious changes
can be applied to generic  $K$-matrix model. For the analogues of (3.25),
(3.28) it gives

\beq
\new
\begin{array}{c}
\tau ^{(K)}_n(t,\bar t) = \left[ \prod  _i
e^{\tilde \varphi _i(s,\bar t)}\right]
{\det _{(ab)}\phi ^{(K,n)}_a(m_b)\over \Delta (m)} = \\
= \left[ \prod  _i e^{\tilde \varphi _i(s,\bar t)}\right]
{\det _{(ab)}\bar \phi ^{(K,n)}_a(\lambda _b)\over \Delta (\lambda )}\hbox{ , }
\end{array}
\eeq
where

\beq
\phi ^{(K,n)}_a(m) = m^{-n} \tilde P^{(K)}_{n+a-1}(m)\hbox{ , }
\eeq

\beq
<\tilde P^{(K)}_i, \tilde {\bar P}
^{(K)}_j> = \int   \tilde P^{(K)}_i(h) \tilde {\bar P}
^{(K)}_j(\bar h)e^{h\bar h-\tilde V(h)-\bar V(\bar h)}d\mu ^{(K)}(h,\bar h) =
e^{\tilde \varphi _i(s,\bar t)}\delta _{ij}\hbox{ , }
\eeq

\beq
\bar \phi ^{(K,n)}_a(\lambda ) = \lambda ^{-n} \tilde {\bar Q}
^{(K)}_{n+a-1}(\lambda )\hbox{ , }
\eeq

\beq
<\tilde Q^{(K)}_i,\tilde {\bar Q}
^{(K)}_j> = \int   \tilde Q^{(K)}_i(h) \tilde {\bar Q}
^{(K)}_j(\bar h)e^{h\bar h-V(h)- \tilde {\bar V}
(\bar h)}d\mu ^{(K)}(h,\bar h) = e^{\tilde \varphi _i(t,\bar s)}
\delta _{ij}\hbox{ . }
\eeq

\subsection{Hermitean one-matrix model}

In the previous section we demonstrated the manifest connection between
$\tau $-functions of discrete matrix models and KP $\tau $-function in Miwa
variables. However, in order to be representable as GKM they still need to
arise in a somewhat specific form. Namely, components of the vector
$\{\phi _i(m)\}$ should possess a representation as ``averages",

$$
\phi _i(m) = \left< x^{i-1} \right> _m.
$$
Since in the study of discrete matrix models  $\phi _i(m)$  arise as orthogonal
polynomials  $P_{n+i}(m)$, what is necessary is a kind of integral
representation of these polynomials, with $i$-dependence coming only from the
$x^{i-1}$--factor in the integrand. It is an interesting problem to find out
such kind of representation for various discrete models, but it is easily
available only whenever orthogonal polynomials are associated with the Gaussian
measure: the relevant Hermit polynomials are known to possess integral
representation, which is exactly of the form which we need. We saw that the
choice of the measure is rather arbitrary in the study of discrete matrix
models, as soon as the complicated time-dependence is eliminated from the
measure with the help of Miwa transformation. Therefore, it may be possible to
make this measure Gaussian. The obstacle can arise either in multi-matrix case,
if we do not want to apply Miwa transformation to ``intermediate" times, or in
the interesting version of two-matrix model with one potential fixed [17,20],
or
in the models, involving non-Hermitean matrices. Each of these situations
deserves special analysis. No more details are necessary in the simplest case
of one-matrix Hermitean model, to be discussed below for illustrative purposes.

The statement to be discussed is that Hermitean one-matrix model with the
matrix
size $n$ is equivalent to GKM with $\hat V(X) = X^2/2 - n\log X$. We can
come to this conclusion in various ways.

First of all, we can look at the complete set of Virasoro constraints for this
model, following from the Ward-identity (2.41). It has been derived in [5] and
shown to coincide with the set of Virasoro constraints for Hermitean one-matrix
model, as discovered in [6,16,21].

Second, we can prove that such GKM corresponds to a {\it forced}
Toda-{\it chain} hierarchy. This statement, if combined with the string
equation (2.38), would also lead to the same conclusion.

Third, we can just verify the explicit identity between Gaussian matrix
integrals,

\beq
{\int dH_{n\times n}\det(I-H/M) e^{-SpH^2/2}\over
\int dH_{n\times n} e^{-SpH^2/2}}  =  {\int
dX_{N\times N}\det (I-iX/M)^ne^{-TrX^2/2}\over
\int dX_{N\times N}e^{-TrX^2/2}}\hbox{ .}
\eeq
Note that the size of matrix in the $l.h.s.$ is $n\times n$ and in the $r.h.s.$
is $N\times N$, and these parameters are absolutely independent. This identity
is indeed true for any  $n$  and  $N$, as follows from the reasoning of the
previous subsection 3.1, and integral representation of Hermit polynomials
$\phi _j(\mu ) = He_j(i\mu )$.

Remarkably, we could use this explicit proof (of eq.(3.39)) as an manifest
verification of the commonly accepted belief that either ({\it i}) the string
equation $(L_{-1}$-constraint) plus the fact that it is imposed on
appropriately reduced Toda lattice $\tau $-function or ({\it ii}) the entire
bunch of Virasoro (or $W$-, if necessary) constraints (without any a priori
information about integrable structure) defines the partition function uniquely
(and, in particular, predetermines the validness of all other $W$-constraints,
suitable to the given reduction, in the case ({\it i}), or guarantees that
partition function is, in fact, the appropriate Toda lattice $\tau $-function
in the case ({\it ii})). This belief is, of course, implicit in the first two
of the above ``derivations" of identity between Hermitean one-matrix model and
GKM with $\hat V(X) = X^2/2 - n\log X$.

In the remaining part of this subsection we shall prove that partition function
of such GKM is indeed a $\tau $-function of {\it forced} Toda {\it chain}
reduction of Toda lattice hierarchy.

Let us begin with the word ``forced". Conceptually, the relevant hierarchy
should be forced just because we deal with discrete matrix models, orthogonal
polynomials, and, thus, all $\phi _i(m)$ for $i>0$ should be {\it polynomials}.
It is, however, somewhat more complicated to prove the property (1.5). It even
looks a bit intriguing, since partition function of GKM (the integral at the
$r.h.s.$ of (3.39)) does not seem to vanish for negative  $n$. The resolution
of the puzzle comes from the study of $n$-dependence of the coefficient of
proportionality.

To do this, let us accurately restore all normalizations in $\tau $-function
and use formula (3.39):

\beq
\tau _n = {1\over n!Vol_{U(n)}} \int dH_{n\times n}
e^{-TrV(H)} = {(2\pi )^{n^2/2}\over n!Vol_{U(n)}}{
\int dX_{N\times N}\det (I-X/M) e^{-TrX^2/2}\over
\int dX_{N\times N}e^{-TrX^2/2}}\hbox{ , }
\eeq
where $Vol_{U(n)}$ denotes the volume of unitary group (it appears due to
necessity of integration over angular variables),

\beq
Vol_{U(n)}^{-1}= (2\pi )^{-n(n-1)/2} \prod ^n_{k=1}k!\hbox{ . }
\eeq
This formula can be obtained, for example, by immediate calculation of matrix
Gaussian integral  $I = \int   dH\ e^{-TrH^2} = (2\pi )^{n^2/2}$ and comparing
it with one calculated in orthogonal polynomials technique:  $I = Vol_{U(n)}
\int \prod    dm_i e^{-m^2_i}
\Delta ^2(m) = Vol_{U(n)} n! \prod ^{n-1}_{k=1}$({\it norm of $He_k)  =
Vol_{U(n)}(2\pi )^{n/2} \prod ^n_{k=1} k!$}.
Now we can continue it to negative $n$ using
the formula (see for example [22]):

\beq
\hbox{if   } f(p) = \prod ^p_{i=1} \phi (i)\hbox{ ,   then  } f(-p) =
\prod ^{p-1}_{i=0} \phi (-i)^{-1}\hbox{ . }
\eeq
It gives in our case

\beq
n!Vol_{U(n)}
\stackreb{n<0}{\rightarrow }
(2\pi )^{n(n-1)/2} \prod ^{|n|}_{i=0} \Gamma (-i)\hbox{ , }
\eeq
what leads to the statement (1.5) due to singularities of $\Gamma $-function at
negative integers.

Our next point is to prove that the GKM integral under consideration is indeed
a Toda {\it chain $\tau $}-function, or, what is the same,  $H_{ij} =
{\cal H}_{i-j}$. We neglect, for a moment, the negative-time variables. To
calculate, we use the following Ward identity for the average with potential
$V(X) = X^2/2$ (the logarithmic term is absorbed into the subscript of
$\phi ):$

\beq
\langle x^i\rangle _z = \langle x^{i-1}z\rangle _z +
(i-1)\langle x^{i-2}\rangle _z\hbox{ .}
\eeq
As  $\displaystyle{H_{ij} = \left< \left< {\phi _i(z) \over z^j} \right>
\right> }$ and  $\phi _i(z) = \langle x^{i-1}\rangle _z$ , one
can write down:

$$
\new
\begin{array}{c}
H_{2j} = H_{1,j-1}\hbox{ ,} \\
H_{3j} = H_{2,j-2} + H_{1,j}\hbox{ ,}\\
H_{4j} = H_{3,j-1} + 2H_{2,j}\hbox{ ,} \\
H_{5j} = H_{4,j-1} + 3H_{3,j}\hbox{ ,}   \\
\ldots \hbox{  . }
\end{array}
$$
Then, using the admissible triangle transformation

$$
\new
\begin{array}{c}
\tilde H_{1j} \equiv  H_{1j}\hbox{ ,}\\
\tilde H_{2j} \equiv  H_{2j}\hbox{ ,}  \\
\tilde H_{3j} \equiv  H_{3j} - H_{1j}\hbox{ ,}\\
\tilde H_{4j} \equiv  H_{4j} - 2H_{2j}\hbox{ ,} \\
\tilde H_{5j} \equiv  H_{5j} - 3H_{3j}\hbox{ ,}
\ldots \hbox{  , }
\end{array}
$$
$i.e.$

\beq
\tilde H_{ij} \equiv  \sum  _k A_{ik}H_{kj}
\eeq
with lower-triangle matrix

$$
A = \left(
\begin{array}{cccccc}
1&&&&&\\&1&&&&\\-1&&1&&&\vdots \\&-2&&1&&\\&&-3&&1&\\&&&\ldots &&
\end{array}
\right) \hbox{ ,}
$$
one obtains the property  $\tilde H_{ij} = {\cal H}_{i-j}$. It is evident that
switching on the negative times as well as logarithmic term modifies Ward
identity (3.44) by contributions of smaller powers in $x$ and can be also
absorbed by proper lower-triangle transformation. It completes the proof.

\section{Double-scaling continuum limit in terms of GKM}
\setcounter{equation}{0}

We demonstrated in the previous sections that the discrete Hermitean one-matrix
model is equivalent to GKM with  $\hat V(X) = X^2/2 + n\log X$, while from
[1] (see also very instructive review [23]) and [12,14,15,24] we know that its
double-scaling continuum limit is described by GKM with  $V(X) = X^3/3$. Thus,
we conclude that

\beq
\lim_{d.s.\hbox{ }n\rightarrow \infty }Z_{\{\hat V\}} = Z^2_{\{V\}}.
\eeq
This relation should certainly be understandable just in terms of GKM itself.
Moreover, since GKM is not sensitive to the size of the matrices $N$, it needs
to become just a simple relation between finite dimensional integrals.

In this section we present some ideas about this relation, without going into
details concerning its compatibility with conventional description of
double-scaling limit [7].

To begin with, let us recall that double-scaling continuum limit for the model
of interest implies that only even times
$\displaystyle{t_{2k} = {1\over 2k} Tr
{1\over M^{2k}}}$  should remain non-zero, while all odd times  $t_{2k+1} = 0$.
This obviously implies that the matrix $M$ should be of block form:

\beq
M = \left(
\begin{array}{cc}
{\cal M} & 0\\0 & -{\cal M}
\end{array}\right)
\eeq
and, therefore, the matrix integration variable is also naturally decomposed
into block form:

\beq
X = \left(
\begin{array}{cc}
{\cal X} & {\cal Z}\\{\cal Z} & {\cal Y}
\end{array}
\right) .
\eeq
Then

\beq
\new
\begin{array}{c}
Z_{\{\hat V=X^2/2-n\log X\}} =\\
= \int d{\cal X}d{\cal Y}d^2{\cal Z}\ \det ({\cal X}{\cal Y}-\bar {\cal Z}
{1\over {\cal Y}}{\cal Z}{\cal Y})^n
e^{-Tr\{|{\cal Z}|^2+{\cal X}^2/2+{\cal Y}^2/2-{\cal M}{\cal X}+
{\cal M}{\cal Y}\}}.
\end{array}
\eeq
To take the limit $n\rightarrow \infty $, one should assume certain scaling
behaviour of ${\cal X}$, ${\cal Y}$ and ${\cal Z}$. Moreover,
the notion of {\it double}-scaling
limit implies a specific {\it fine tuning} of this scaling behaviour. So we
shall take

\beq
\new
\begin{array}{c}
{\cal X} = \alpha (i\beta I + x), \\
{\cal Y} = \alpha (-i\beta I + y),      \\
{\cal Z} = \alpha \zeta , \\
{\cal M} = \alpha ^{-1}(i\gamma I + m)
\end{array}
\eeq
with some large real $\alpha $, $\beta $ and $\gamma $. If expressed through
these variables, the action becomes:

\beq
\new
\begin{array}{c}
Tr\{|{\cal Z}|^2 + {\cal X}^2/2 + {\cal Y}^2/2 - {\cal M}{\cal X} +
{\cal M}{\cal Y} - n\log ({\cal X}{\cal Y} - \bar {\cal Z}{1\over
{\cal Y}}{\cal Z}{\cal Y})\} = \\
= {\gamma ^2\over 2}Tr\{(i\beta I + x)^2 + {\gamma ^2\over 2}Tr(i\beta I -yx)^2
+ \gamma ^2|z|^2\} - Tr(i\alpha I + m)(2i\beta I + x - y) - \\
- nTr \log \ \beta ^2\gamma ^2\{1 - i {x-y\over \beta } +
{xy\over \beta ^2} - {|\zeta |^2\over \beta ^2}(1 + o(1/\beta ))\} =
\end{array}
\eeq
$$
= [2\alpha \beta  - \beta ^2\gamma ^2 - 2n\ \log \ \beta \gamma ] Tr\ I -
2i\beta \ Tr\ m +
\eqno{(A)}
$$
$$
+ i(\beta \gamma ^2 - \alpha  + {n\over \beta })(Tr\ x - Tr\ y) +
{1\over 2}(\gamma ^2 - {n\over \beta ^2})(Tr\ x^2 + Tr\ y^2) +
\eqno{(B)}
$$
$$
+ (\gamma ^2 - {n\over \beta ^2}) Tr |\zeta |^2 -
\eqno{(C)}
$$
$$
- Tr mx + Tr\ my + {in\over 3\beta ^2}Tr(x^3 - y^3) +
\eqno{(D)}
$$
$$
+ {\cal O}(n/\beta ^4) + {\cal O}(|\zeta |^2 {n\over \beta ^3}).
\eqno{(E)}
$$
We want to adjust the scaling behaviour of $\alpha $, $\beta $ and $\gamma $ in
such a way that only the terms in the line $(D)$ survive. This goal is achieved
in several steps.

The line ({\it A}) describes normalization of functional integral, it does not
contain $x$ and $y$. Thus, it is not of interest for us at the moment.

Two terms in the line $(B)$ are eliminated by adjustment of $\alpha $ and
$\gamma $:

\beq
\gamma ^2 = {n\over \beta ^2}\hbox{ , }  \alpha  = {2n\over \beta }\hbox{ .}
\eeq
As we shall see soon,  $\gamma ^2 = n/\beta ^2$ is large in the limit of
$n\rightarrow \infty $ . Thus, the term $(C)$ implies that the fluctuations of
$\zeta $-field are severely suppressed, and this is what makes the terms of the
second type in the line $(E)$ negligible. More general, this is the reason for
the integral  $Z_{\{\hat V\}}$
to split into a product of two independent integrals leading to the square
of partition function in the limit $n\rightarrow \infty $ (this splitting is
evident as, if ${\cal Z}$ can be neglected, the only mixing term
$\displaystyle{\log \det
\left(
\begin{array}{cc}
{\cal X} & {\cal Z}\\{\cal Z} & {\cal Y}
\end{array}
\right)} $  turns into  $\log {\cal X}{\cal Y} = \log {\cal X} +
\log {\cal Y}$).

Thus, we remain with a single free parameter $\beta $ which can be adjusted so
that

\beq
{\beta ^3\over n}\rightarrow const\ \ \
\hbox{   as }\ \ \ n\rightarrow \infty \ \ \
(i.e\hbox{. }  \beta  \sim  \sqrt{n}  ),
\eeq
making the terms in the last line $(E)$ vanishing and the third term in the
line $(D)$ finite.

This proves the statement (4.1) in a rather straightforward manner, without
addressing directly to the complicated matter like Kazakov's change of time
variables, reformulation of Virasoro constraints and so on [7]. We do not go
into more details here, but point out one important detail. That is, the
possibility to eliminate all the terms of original potential  $\hat V_+(X)$ of
degree $K=2$ by the contributions coming from expansion of logarithm in such a
way that the $(K+1)$-th power of expansion survives is due to careful fine
tuning of parameters of original  $\hat V_+(X)$. This is just the idea involved
in the notion of double-scaling limit and for higher-degree potentials it
should be replaced by ``$K$-scaling" limit which turns  $Z_{\{V=X^K-
n\log X\}}$ into $Z^2_{\{V=X^{K+1}\}}$ as $n\rightarrow \infty $.

\section{CONCLUSION}

To conclude, let us first remind the main idea of the paper. The proposed
Generalized Kontsevich Model seems now to be close to a unified theory of all
(discrete and continuous) matrix models. By introducing of ``negative-" and
``zero-" time variables it results to be a $\tau $-function of the Toda lattice
hierarchy and allows one to unify all matrix models in this framework. In
particular, the original Hermitean one-matrix model is, in these terms, the GKM
with ``trivial" potential  $X^2$ (non-trivial models start from original
Kontsevich's one with  $X^3$ potential) but with non-vanishing zero-time. This
must give an opportunity to study all subtle questions in this unified
language (the continuum limit of discrete models among them -- the sketch of
such treatment was proposed in the sect.4). Besides, the introducing of
zero-time variable leads to the notion of ``natural basis" in Grassmannian with
property  $\phi ^{(n)}_i(z) = z^{-n} \phi ^{(0)}_{i+n}(z)$. This basis differs
from the canonical one (which is singled out in the Segal--Wilson construction
[27] and corresponds to standard representation of $\phi _i$ through the
fermionic averages in spirit of GKM [1]) and coincides with one inherited from
GKM integral (2.25).

One of the most important items is the matter of the sect.2.3 concerning with
(generalized) string equation. We see now that the string equation does not
determine the point of infinite-dimensional Grassmannian uniquely -- it allows
one to deform it in accordance with negative-time Toda flows. In this sense,
the introducing of the negative times would rather correspond to handle-gluing
operators being added to action of string model, while the positive times to
the motion in the space of string models. What is really spoilt by these
negative times is the particular {\it reduction} condition, which is
important in order to determine a particular string model with
$\displaystyle{c = 1 -
6{(q-q')^2\over qq'}}$. This condition is
also spoilt by perturbation of the pure
monomial potentials  $X^{K+1}$ by lower order irrelevant operators (in the
sense of the universality class defined by the string equation), however the
physical properties of these ``deformed" models should be the same as those of
the original one. In particular, it means that the choice of normalization
of the basis in Grassmannian given by GKM  $\phi _i(z) \sim  \int
dx\ e^{-\hat V(x)+V'(z)x}$ could be more flexible.

We have only touched in this paper the most serious problem of the continuum
limit, which has been done correctly up to now only for the Hermitean
one-matrix
model [7]. As it has been shown in the sect.4, this continuum limit in
principle can be done by purely ``field-theory" technique in the framework of
GKM, though some questions are still not completely clear. It allows us to hope
that the proposed in [1] and above formalism will shed light on the serious
problem of the continuum limit in multi-matrix models, among which the
asymmetric two-matrix model proposed in [17,20] seems to be the most
perspective one.

Finally, we have to repeat that we do not agree with the interpretation of [5]
that the potential  $X^2-n\log X$  describe consistently  $c = 1$  string
theory,
or, equivalently, matrix quantum mechanics . However, we still hope that the
limit  $c \rightarrow  1$  can be found in the framework of GKM and, probably,
by using of non-polynomial potentials.

All these problems deserve further investigation and we hope to return to them
elsewhere.

\bigskip
We are grateful to L.Chekhov, A.Orlov, A.Sagnotti and A.Zabrodin for deep and
very useful discussions.

\section*{Appendix A. Integrable hierarchies in the operator formalism}
\def\theequation{A.\arabic{equation}}
\setcounter{equation}{0}

Here we would like to describe some basic ingredients of the integrable
hierarchies in the terms of the massless fermions. This approach was initiated
in the series of papers [25] and it appears to be very fruitful for description
of the general structure of the large variety of nonlinear integrable equations
like KP, Toda etc.

Let us define the fermionic operators on sphere

\beq
\psi (z) =\sum _{k\in {\bf Z}}\psi _kz^k\hbox{ , }  \psi ^\ast (z)
=\sum _{k\in {\bf Z}}\psi ^\ast _kz^{-k} ,
\eeq
where fermionic modes satisfy the usual anti-commutation relations:

\beq
\{\psi _k\hbox{, } \psi ^\ast _m\} = \delta _{km}\hbox{ , }  \{\psi _k\hbox{, }
\psi _m\} = \{\psi ^\ast _k\hbox{, } \psi ^\ast _m\} = 0\hbox{ .}
\eeq
The Dirac vacuum  $|0\rangle $  is defined by the conditions:

\beq
\psi _k|0\rangle  = 0\ ,\ k < 0\ ;\ \psi ^\ast _k|0\rangle  = 0\ ,\ k \geq
0\hbox{  .}
\eeq
We also need to introduce the ``shifted" vacua constructed from  $|0\rangle $
as follows

\beq
|n\rangle  =  \left\{
\begin{array}{l}
\psi _{n-1}\ldots \psi _0 \left|  0\rangle ,\ \ \ n\geq 0 \right. \\
\psi ^\ast _n
\ldots \psi ^\ast _{-1}\left| 0\rangle ,\ \ \  n <  0 \right.
\end{array}
\right.
\eeq
and satisfying the obvious conditions

\beq
\psi _k|n\rangle  = 0\ ,\ k < n\hbox{ ; }   \psi ^\ast _k|n\rangle  = 0\ ,\ k
\geq  n .
\eeq
{}From fermionic modes one can built the  $U(1)$--currents

\beq
J_k =\sum _{i\in {\bf Z}}\psi _i\psi ^\ast _{i+k}\hbox{ , }  J_{-k} \equiv
\bar J_k\hbox{ , }  k \in  {\bf Z}_+
\eeq
and define ``Hamiltonians"

\beq
H(x) = \sum ^\infty _{k=1}x_kJ_k\hbox{ , }  \bar H(x) =
\sum ^\infty _{k=1}y_k\bar J_k\hbox{ ,}
\eeq
where  $\{x_k\}$  and  $\{y_k\}$  are half-infinite sets of independent time
variables (``positive" and ``negative" times correspondingly; in main text
$\{x_k\} \equiv  \{t_k\}$, $\{y_k\} \equiv  \{-t_{-k}\}$, $k > 0$) which
generate the evolution of nonlinear system.

Let  $g$  be an arbitrary element of the Clifford group which does not mixes
the $\psi$- and $\psi ^\ast $- modes :

\beq
g = :\exp [\sum     A_{km}\psi _k\psi ^\ast _m]: ,
\eeq
where  :  :  denotes the normal ordering with respect to the Dirac vacuum
$|0\rangle $. Then it is well known [10,25] that

\beq
\tau _n(x,y) = \langle n|e^{H(x)}ge^{-\bar H(y)}|n\rangle
\eeq
solves the two-dimensional Toda lattice hierarchy, $i.e$. is the solution to
the whole set of the Hirota bilinear equations. Any particular solution depends
only on the choice of the element  $g$ (or, equivalently it can be uniquely
described by the matrix  $A_{km}$). From the eqs.(A.2) one can conclude that
any element in the form (A.8) rotates the fermionic modes as follows

\beq
g\psi _kg^{-1} = \psi _jR_{jk}\hbox{ , }  g\psi ^\ast _kg^{-1} =
\psi ^\ast _jR^{-1}_{kj} ,
\eeq
where the matrix  $R_{jk} $ can be expressed through  $A_{jk}$ (see [26]). We
will see below that the general solution (A.9) can be expressed in the
determinant form with explicit dependence of  $R_{jk}$ . In order to calculate
$\tau $-function  we need some more notations. Using commutation relations
(A.2) one can obtain the evolution of  $\psi (z)$ and  $\psi ^\ast (z)$ in
times $\{x_k\}$, $\{y_k\}$ in the form

\beq
\psi (z,x) \equiv  e^{H(x)}\psi (z)e^{-H(x)} = e^{\xi (x,z)}\psi (z)\hbox{  ,}
\eeq

\beq
\psi ^\ast (z,x) \equiv  e^{H(x)}\psi ^\ast (z)e^{-H(x)} =
e^{-\xi (x,z)}\psi ^\ast (z)\hbox{  ;}
\eeq

\beq
\psi (z,\bar y) \equiv  e^{\bar H(y)}\psi (z)e^{-\bar H(y)} =
e^{\xi (y,z^{-1})}\psi (z)\hbox{ , }
\eeq

\beq
\psi ^\ast (z,\bar y) \equiv  e^{\bar H(y)}\psi ^\ast (z)e^{-\bar H(y)} =
e^{-\xi (y,z^{-1})}\psi (z)\hbox{ ,}
\eeq
where

\beq
\xi (x,z) = \sum ^\infty _{k=1}x_kz^k\hbox{ .}
\eeq
Let us define the the Schur polynomials  $P_k(x)$:

\beq
{\cal P}[z|x] = e^{\xi (x,z)} = \sum ^\infty _{k=0}P_k(x)z^k\hbox{ ;}
\eeq
then from eqs. (A.11)-(A.14) one can easily obtain the evolution of the
fermionic modes:

\beq
\psi _k(x) \equiv e^{H(x)}\psi _ke^{-H(x)}
=\sum ^{\infty}_{m=0} \psi _{k-m}P_m(x) ,
\eeq

\beq
\psi ^{\ast }_k(x) \equiv e^{H(x)}\psi _k^{\ast }e^{H(x)} =
\sum ^{\infty}_{m=0} \psi ^{\ast }_{k+m}P_m(-x) ;
\eeq

\beq
\psi _k(\bar y) \equiv e^{\bar H(y)}\psi
_ke^{-\bar H(y)} =\sum ^{\infty} _{m=0} \psi
_{k+m}P_m(y);
\eeq

\beq
\psi _k^{\ast }(\bar y) \equiv e^{\bar H(y)}\psi
_k^{\ast }e^{-\bar H(y)}=\sum ^{\infty} _{m=0} \psi
_{k-m}^{\ast } P_m(-y).
\eeq
It is useful to introduce the totally occupied state  $|-\infty \rangle $  (see
definition (A.4) in the limit $n \rightarrow  -\infty $) which satisfies the
requirements

\beq
\psi ^\ast _i|-\infty \rangle  = 0\hbox{  , }  i \in  {\bf Z}\hbox{  .}
\eeq
Then any shifted vacuum can be generated from this state as follows:

\beq
|n\rangle  = \psi _{n-1}\psi _{n-2} ...|-\infty \rangle \hbox{  .}
\eeq
Note that the action of an any element  $g$  of the Clifford group (and, as
consequence, the action of $e^{-\bar H(y)} $) on  $|-\infty \rangle $  is very
simple: $g|-\infty \rangle  \sim  |-\infty \rangle $, so using (A.18) and
(A.19) one can obtain from eq.(A.9):

\beq
\new
\begin{array}{c}
\tau _n(x,y) =
\langle -\infty |...\psi ^\ast _{n-2}(-x)\psi ^\ast _{n-1}(-x)g\psi _{n-1}(-%
\bar y)\psi _{n-2}(-\bar y) ...|-\infty \rangle  \sim \\
\sim \det [\langle -\infty |\psi ^\ast _i(-x)g\psi _j(-\bar y)g^{-1}|-\infty
\rangle ]\left| _{i,j \leq  n-1}\right.\hbox{ .}
\end{array}
\eeq
Using (A.10) it is easy to see that

\beq
g\psi _j(-\bar y)g^{-1} =
\sum _{m,k} P_m(-y)\psi _kR_{k,j+m}
\eeq
and the ``explicit" solution of the two-dimensional Toda lattice has the
determinant representation:

\beq
\tau _n(x,y) \sim \left. \det \ {\hat H}_{i+n,j+n} (x,y)
\right| _{i,j<0}\hbox{ ,}
\eeq
where

\beq
{\hat H}_{ij}(x,y) = \sum _{k,m} R_{km}P_{k-i}(x)P_{m-j}(-y)\hbox{  .}
\eeq
The ordinary solutions to KP hierarchy [25] correspond to the case when the
whole evolution depends only of positive times $\{x_k\}$; negative times
$\{y_k\}$ serve as parameters which parameterize the family of points in
Grassmannian and can be absorbed by re-definition of the matrix $R_{km}$. Then
$\tau $-function of (modified) KP hierarchy has the form

\beq
\tau _n(x) = \langle n|e^{H(x)}g(y)|n\rangle \sim \det [\sum  _k
R_{k,j+n}(y)P_{k-i-n}(x)]\left| _{i,j<0}\right.\hbox{ ,}
\eeq
where  $g(y) \equiv  ge^{-\bar H(y)}$ and

\beq
R_{kj}(y) \equiv  \sum  _m R_{km}P_{m-j}(-y)\hbox{  .}
\eeq
One can consider the reduction to the Toda chain hierarchy after imposing the
condition on the element  $g$ [10]

\beq
[J_k + \bar J_k,g] = 0
\eeq
which is equivalent to constraint

\beq
[\Lambda  + \Lambda ^{-1},R] = 0\hbox{  .}
\eeq
In this case

$$
ge^{-y_k\bar J_k} = e^{-y_k\bar J_k} e^{-y_kJ_k} g\ e^{y_kJ_k}
$$
and $\tau $-function depends (up to the trivial factor) only on times  $\{x_k -
y_k\}$:

\beq
\tau _n(x,y) = e^{\sum kx_ky_k} \langle n|e^{H(x-y)}g|n\rangle \hbox{  .}
\eeq
The reduction (A.30) has an important
solution\footnote{Generally the solutions
$R_{nk}=R_{n+k}$ and  $R_{nk}=R_{n-k}$ are
different, but for forced hierarchy, when $\tau $-function is the determinant
of finite matrix, these two solutions are equivalent due to possibility
to reflect
matrix with respect to vertical axis without changing the determinant.}

\beq
R_{km} = R_{k+m}\hbox{ .}
\eeq
In this case the matrix ${\hat H}_{ij}$ defined by eq.(A.26) evidently
satisfies
the
relations ${\hat H}_{ij} = {\hat H}_{i+j}$ and

\beq
(\partial _{x_k}+ \partial _{y_k}){\hat H}_{i+j} = 0
\hbox{     for any }\ \ k<n-i
\ ,\ k<n-j
\eeq
due to the properties of Schur polynomials

\beq
\partial _{x_n}P_k(x) = P_{k-n}(x).
\eeq
The property (A.33) certainly does not imply that the corresponding
$\tau $-function depends only on difference of times because of restriction of
values of $k$, but it restores correct dependence of times with taking into
account of exponential in (A.31).

Now let us establish the corespondence between matrices $R_{ij}$ (eq.(A.10)),
${\hat H}_{ij}$ (eq.(A.26)) and $T_{ij}$ (eq.(2.12)), $H_{ij}$ (eq.(2.2)).
These equations transform into each other under charge conjugation of
fermions $\psi_k \to \psi_{-k-1}^{\ast}$ ($i.e.$
vacua transform as follows:
$|n \rangle \to |-n \rangle$) and correspondence: $T_{ij}=R^{-1}_{j-1,-i-1}$
and $\hat H_{ij}=H_{ij}$.

\section*{Appendix B.  Fermionic representation of the matrix models}
\def\theequation{B.\arabic{equation}}
\setcounter{equation}{0}

Here we would like to discuss the fermionic language for (multi-) matrix
models.
Namely, we shall show that the $\tau $-function of the two-dimensional Toda
lattice (A.9) describes the whole variety of the (multi-) matrix models for
some specific choice of the element  $g$. Since one should reproduce the
partition function of the {\it matrix} models from eq.(A.9), we shall deal with
the forced hierarchies (see (1.5) and discussion in the sect.2.1), $i.e.$

\beq
\tau _n = 0\hbox{  , }  n < 0\hbox{  .}
\eeq
Therefore, it is reasonable to consider the point of the Grassmannian in the
form

\beq
g = g_0P_+ ,
\eeq
where  $P_+$ is the projector onto positive states:

\beq
P_+|n\rangle  = \theta (n)|n\rangle .
\eeq
There is exist a natural fermionic projector

\beq
P_+ = :\exp [\sum _{i<0}\psi _i\psi ^\ast _i]:
\eeq
with the properties

\beq
P_+\psi ^\ast _{-k} = \psi _{-k}P_+ = 0\hbox{  , }  k > 0\ ;
\eeq

\beq
\left[ P_+,\psi _k\right] = [P_+,\psi ^{\ast} _k] = 0\hbox{  , } k \geq  0\ ;
\eeq

\beq
P^2_+ = P_+ .
\eeq
The insertion of such projector into eq.(A.9) naturally leads us to conclusion
that  $g_0$ should depends only on $\psi _k$ and $\psi ^\ast _k$ with $k \geq
0$. We use the choice

\beq
g_0 = :\exp \left\{ \left(\int _\gamma  A(z,w)\psi _+(z)
\psi ^\ast _+(w^{-1})dzdw \right) - \sum _{i\geq 0} \psi _i\psi ^\ast _i
\right\}:\ \ ;
\eeq
where $\psi _+(z) =\sum _{k\geq 0}\psi _kz^k$ , $\psi ^\ast _+(z)
=\sum _{k\geq 0}\psi ^\ast _kz^{-k}$ and  $\gamma $  is some contour of
integration. In what follows we shall also use the projector

\beq
P_- = :\exp [- \sum _{i\geq 0}\psi _i\psi ^\ast _i]:
\eeq
with the properties

\beq
P_-\psi _k = \psi _k^\ast P_- = 0\hbox{  , }  k \geq 0\  ;
\eeq

\beq
\left[ P_-,\psi _{-k}\right] = [P_-,\psi _{-k}{^\ast} ] = 0\hbox{  , }
k > 0\ ;
\eeq

\beq
P^2_- = P_- .
\eeq
Now one should calculate the state

\beq
g_0P_+e^{-\bar H(y)}|n\rangle \ .
\eeq
It is easy to see that this state vanishes when $n < 0$. Indeed, using
eqs.(A.4) and (A.20) one can obtain that with $n < 0$ the state

$$
e^{-\bar H(y)}|n\rangle  = \psi ^\ast _{-n}(-\bar y)\hbox{ ... }
\psi ^\ast _{-1}(-\bar y)e^{-\bar H(y)}|0\rangle
$$
contains only negative modes $\psi ^\ast _{-m} ( m > 0)$. Therefore the action
of $P_+$ annihilates this state due to eq.(B.5). For $n \geq  0$ using
eqs.(A.4), (A.19) and (B.6) we have

\beq
P_+e^{-\bar H(y)}|n\rangle  = \psi _{n-1}(-\bar y)\hbox{ ... }
\psi _0(-\bar y)P_+e^{-\bar H(y)}|0\rangle .
\eeq
Then we use the fact that

\beq
P_+e^{-\bar H(y)}|0\rangle  = |0\rangle \hbox{  .}
\eeq
\underline {Proof of eq.(B.15).} Let us denote

$$
|y\rangle  = P_+e^{-\bar H(y)}|0\rangle \hbox{  .}
$$
Then

$$
{\partial \over \partial y_k} |y\rangle  = P_+e^{-\bar H(y)} \sum ^{k-1}_{i=0}
\psi ^{\ast} _{i-k}\psi _i|0\rangle  = 0
$$
due to eqs.(A.3) and (B.5). Since $\displaystyle{|y\rangle \left|
_{y_k=0} \right.= |0\rangle}$, the eq.(B.15) is proved.

Therefore, we have

\beq
\new
\begin{array}{c}
g_0P_+e^{-\bar H(y)}|n\rangle  = g\ \psi (-\bar y)\ldots
\psi (-\bar y)|0\rangle  = \\
= \sum     {1\over m!} \int _\gamma  \prod ^m_{i=1}A(z_i,w_i)dz_idw_i
\psi _+(z_1)\ldots\psi _+(z_m) \times \\
\times P_-
\psi ^\ast _+(w^{-1}_m)\ldots\psi ^\ast _+(w^{-1}_1)\psi _{n-1}(-\bar y)
\ldots \psi _0(-\bar y)|0\rangle
\end{array}
\eeq
with using of the eqs.(B.8) and (B.9). Now we shall see that only the term with
$m = n$ gives a non-zero contribution in the infinite sum (B.16). Indeed, for
$m > n$ the state
$\psi ^\ast _+(w^{-1}_m)\ldots\psi ^\ast _+(w^{-1}_1)\psi _{n-1}(-\bar y)
\ldots \psi _0(-\bar y)|0\rangle $
vanishes, because in this case some positive modes
in $\psi ^\ast _+(w^{-1}_i)$ will reach the vacuum $|0\rangle $ and annihilate
it. Vice versa, for $m <n$ some positive modes in $\psi _k(-\bar y)$ will reach
the projector $P_-$ and due to eq.(B.10) it is zero. Therefore,

\beq
\new
\begin{array}{c}
g_0P_+e^{-\bar H(y)}|n\rangle
= {1\over n!} \int _\gamma  \prod ^n_{i=1}A(z_i,w_i)dz_idw_i
\psi _+(z_1)\ldots \psi _+(z_n) \times \\
\times P_-
\psi ^\ast _+(w^{-1}_n)\ldots \psi ^\ast _+(w^{-1}_1)\psi _{n-1}(-\bar y)
\ldots  \psi _0(-\bar y)|0\rangle .
\end{array}
\eeq
Now we use the following proposition:

\beq
\new
\begin{array}{c}
\psi ^\ast _+(w^{-1}_n)\ldots \psi ^\ast _+(w^{-1}_1)\psi _{n-1}(-\bar y)
\ldots \psi _0(-\bar y)|0\rangle  = \\
= \Delta (w)\exp [-\sum ^n_{j=1}\xi (y,w_j)]|0\rangle \hbox{  .}
\end{array}
\eeq
\underline {Proof.}
Since the number of creation ($w.r.t.$ to $|0\rangle $) operators
$\psi _i(-\bar y)$ equals to the number of annihilation operators
$\psi ^\ast _+(w^{-1}_j)$, then it is obvious that after normal re-ordering

$$
\psi ^\ast _+(w^{-1}_n)...\psi ^\ast _+(w^{-1}_1)\psi _{n-1}(-\bar y)\ldots
\psi _0(-\bar y)|0\rangle  = const\cdot |0\rangle
$$
and, consequently,

$$const =
\langle 0|\psi ^\ast _+(w^{-1}_n)\ldots \psi ^\ast _+
(w^{-1}_1)\psi _{n-1}(-\bar y)
\ldots \psi _0(-\bar y)|0\rangle  =
$$
$$
= \det [\langle 0|\psi ^\ast _+(w^{-1}_i)\psi _{j-1}(-\bar y)|0\rangle ]
\left| _{i,j = 1,\ldots ,n}\right.
$$
and using eqs.(A.20), (A.16) one can obtain

$$
\langle 0|\psi ^\ast _+(w^{-1}_i)\psi _{j-1}(-\bar y)|0\rangle  = w^{j-1}_i
e^{-\xi (y,w_i)}\hbox{ ;}
$$
thus,

$$const = \det [w^{j-1}_i e^{-\xi (y,w_i)}] =
\Delta (w)\exp [-\sum ^n_{j=1}\xi (y,w_j)]  .
$$
After substitution of eq.(B.18) into eq.(B.17) and using the obvious fact that
$P_-|0\rangle  = |0\rangle $ and $\psi _-(z_i)|0\rangle  = 0$ one can obtain

\beq
\new
\begin{array}{c}
g_0P_+e^{-\bar H(y)}|n\rangle  = \\
= {1\over n!} \int _\gamma  \prod ^n_{i=1}A(z_i,w_i)e^{-\xi (y,w_i)}dz_idw_i
\Delta (w)\psi (z_1)\ldots \psi (z_n)|0\rangle .
\end{array}
\eeq
Using one of the basic formulas [25] (which can be simply proved by
bosonization technique)

\beq
\psi (z_1)\ldots \psi (z_n)|0\rangle  = \Delta (z)
\exp [\bar H(\sum ^n_{i=1}\epsilon (z_i)]|0\rangle ,
\eeq
where $\epsilon (z_i)$ is the vector with components
$\displaystyle{\epsilon _k(z_i) =
{1\over k} z^k_i}$ , we have the desired result:

\beq
\new
\begin{array}{c}
g_0P_+e^{-\bar H(y)}|n\rangle  = \\
= {1\over n!} \int _\gamma  \prod ^n_{i=1}A(z_i,w_i)e^{-\xi (y,w_i)}dz_idw_i
\Delta (w)\Delta (z) \exp [\bar H(\sum ^n_{i=1}\epsilon (z_i)]|0\rangle .
\end{array}
\eeq
Thus, finally we obtain:

\beq
\new
\begin{array}{c}
\tau _n(x,y) = \langle n|e^{H(x)}g_0P_+e^{-\bar H(y)}|n\rangle  = \\
= {1\over n!} \int _\gamma
\Delta (w)\Delta (z)\prod ^n_{i=1}A(z_i,w_i)e^{\xi (x,z_i)-\xi
(y,w_i)}dz_idw_i  .
\end{array}
\eeq
Let us consider some particular cases.

i) $\gamma  = (-\infty ,+\infty )$ , $A(z,w) = \delta (z - w)$. Then one can
recover the case of Hermitean one-matrix model.

ii) $\gamma $ is the small circle around the origin in the complex $z$-plane,
and $A(z,w) = \displaystyle{{1\over 2\pi iz}\delta (z - w^{-1})}$.
Then $g_0 = 1$ and
$\tau _n(x - y)$ is the $\tau $-function for the symmetric unitary model.

iii) $\gamma  = (-\infty ,+\infty )$ , $A(z,w) = e^{zw}$ . This is Hermitean
two-matrix model.

iv) $\gamma  = (-\infty ,+\infty )$ . Let us denote $z_i \equiv  z^{(1)}_i$ ,
$w_i \equiv  z^{(p)}_i$ , $x \equiv  - t^{(1)}$ , $y \equiv  t^{(p)}$ and

\beq
A(z^{(1)}_i,z^{(p)}_i) =  -\int ^{+\infty }_\infty  \prod ^{p-1}_{j=1}
dz^{(j)}_i \exp \{-\sum ^{p-1}_{j=2} \xi (t^{(j)},z^{(j)}_i) +
\sum ^{p-1}_{j=1} z^{(j)}_i z^{(j+1)}_i\}  .
\eeq
Then we obtain the partition function for Hermitean $p$-matrix model which
originates from the matrix integrals

\beq
\int   DZ^{(1)}\ldots
DZ^{(p)}\exp \{- Tr \sum ^{p-1}_{j=1} \xi (t^{(j)},Z^{(j)}) +
Tr \sum ^{p-1}_{j=1} Z^{(j)} Z^{(j+1)}\} .
\eeq
The only point is that in this case  $g$  is parameterized by the set of
additional times $t^{(j)}$, $j = 2,\ldots ,p-1$. One can obtain the
multi-matrix model with time-independent  $g$  in the context of the
multi-component Toda lattice hierarchy.

\bigskip
\underline {Determinant representation.}

Using eqs.(A.4), (B.7) and the fact that $[P_+,g_0] = 0$, we have:

$$
\tau _n(x,y) =
\langle 0|\psi ^\ast _0\ldots
\psi ^\ast _{n-1}e^{H(x)}g_0P_+e^{-\bar H(y)}\psi _{n-1
}\ldots \psi _0|0\rangle  =
$$
$$
=
\langle 0|e^{H(x)}\psi ^\ast _0(-x)\ldots
\psi ^\ast _{n-1}(-x)P_+g_0P_+\psi _{n-1}%
(-\bar y)\ldots \psi _0(-\bar y)e^{-\bar H(y)}|0\rangle \hbox{ .}
$$
Since $\psi ^\ast _i(-x)$ and $\psi _i(-\bar y)$ contain only positive modes
(see eqs.(A.18) and (A.19)), due to (B.6) and (B.15) one can obtain

\beq
\new
\begin{array}{c}
\tau _n(x,y) =
\langle 0|\psi ^\ast _0(-x)\ldots
\psi ^\ast _{n-1}(-x)g_0\psi _{n-1}(-\bar y)\ldots
\psi _0(-\bar y)|0\rangle  = \\
\det [\langle 0|\psi ^\ast _i(-x)g_0\psi _j(-\bar y)|0\rangle ]
\left| _{i,j=0,...,n-1} \right.  .
\end{array}
\eeq
The same arguments used for transition from eq.(B.16) to eq.(B.17) when applied
to eq.(B.25) lead to conclusion that only linear term in $A(z,w)$ contributes,
so using eq.(B.10) we have

\beq
\new
\begin{array}{c}
\langle 0|\psi ^\ast _i(-x)g\psi _j(-\bar y)|0\rangle  = \int _\gamma
A(z,w)dzdw
\langle 0|\psi ^\ast _i(-x)
\psi _+(z)P_-\psi ^\ast _+(w^{-1})\psi _j(-\bar y)|0
\rangle  = \\
= \int _\gamma  z^iw^jA(z,w)e^{\xi (x,z)-\xi (y,w)}dzdw  =
\partial ^i_x (-\partial _y )^j\int _\gamma
A(z,w)e^{\xi (x,z)-\xi (y,w)}dzdw
\end{array}
\eeq
and, finally, one can obtain the expression for $\tau $-function in the
determinant form:

\beq
\tau _n(x,y) = \det [\partial ^i_{x_1}(-\partial _{y_1})^j\int _\gamma
A(z,w)e^{\xi (x,z)-\xi (y,w)}dzdw] \left|
_{i,j=0,...,n-1}\right.\hbox{ .}
\eeq
Again, the consideration of particular choices of $A(z,w)$ (see discussion
below eq.(B.22)) leads to representation of $\tau $-function for Hermitean,
unitary etc. models in the determinant form.

\bigskip
\bigskip
\begin{center}{{\Large\bf References}}\end{center}
\medskip
1. S.Kharchev et al. Unification of all string model with $c<1$, Phys.Lett.
{\bf B} (in press);

\hspace{.5in}
Towards unified theory of 2d gravity, Nucl.Phys. {\bf B} (in press)\\
2. E.Witten  Nucl.Phys., {\bf B340} (1990) 281\\
3. M.Kontsevich  Funk.Anal. \& Prilozh., {\bf 25} (1991) 50\\
4. S.Kharchev et al. Nucl.Phys., {\bf B366} (1991) 569\\
5. L.Chekhov, Yu.Makeenko  A hint on the external field problem for matrix
models,

\hspace{.5in}
Preprint NBI-HE-92-06, January 1992\\
6. A.Gerasimov et al. Nucl.Phys., {\bf B357} (1991) 565\\
7. Yu.Makeenko et al. Nucl.Phys., {\bf B356} (1991) 574\\
8. P.J.Hansen, D.J.Kaup  J.Phys.Soc.Japan, {\bf 54} (1985) 4126\\
9. A.Leznov, M.Saveliev  Physica, {\bf D3} (1981) 256\\
10. K.Ueno, K.Takasaki  Adv.Studies in Pure Math., {\bf 4} (1984) 1\\
11. M.Fukuma, H.Kawai, R.Nakayama  Int.J.Mod.Phys., {\bf A6} (1991) 1385\\
12. A.Marshakov, A.Mironov, A.Morozov  Phys.Lett. $B274$ (1992) 280\\
13. D.J.Gross, M.J.Newman  Phys.Lett., {\bf B266} (1991) 291 \\
14. D.J.Gross, M.J.Newman  Unitary and Hermitean matrices in an external field
II: the

\hspace{.5in}
Kontsevich model and continuum Virasoro constraints, Preprint
PUPT-1282,

\hspace{.5in}
December 1991\\
15. E.Witten  On the Kontsevich model and other models of two dimensinal
gravity,

\hspace{.5in}
Preprint IASSNS$-HEP-91/24$, June 1991\\
16. A.Mironov, A.Morozov  Phys.Lett., {\bf B252} (1990) 47 \\
17. A.Marshakov, A.Mironov, A.Morozov  Pisma $v$ ZhETF, {\bf 54} (1991) 536;

\hspace{.5in}
{}From Virasoro constraints in Kontsevich's model to $W$-constraints in
2-matrix

\hspace{.5in}
models, Preprint $FIAN/TD/05-91$, ITEP$-M-5/91$, October 1991\\
18. E.Brezin, V.Kazakov, Al.B.Zamolodchikov  Nucl.Phys., {\bf B338} (1990) 673

D.J.Gross, N.Miljkovic  Phys.Lett., {\bf B238} (1990) 217

P.Ginsparg, J.Zinn-Justin  Phys.Lett., {\bf B240} (1990) 333

D.J.Gross, I.R.Klebanov  Nucl.Phys., {\bf B344} (1990) 475\\
19. A.Erdelyi et al. (The Bateman manuscript project), Higher transcendetal
functions,

\hspace{.5in}
McGraw-Hill, New York, 1953\\
20. E.Gava, K.Narain  Phys.Lett., {\bf B263} (1991) 213\\
21. J.Ambjorn, J.Jurkevich, Yu.Makeenko  Phys.Lett., {\bf B251} (1990) 517\\
22. Vl.S.Dotsenko  Three-point correlation functions of the minimal conformal
theories

\hspace{.5in}
coupled to 2d gravity, Preprint PAR-LPTHE 91-18, February 1991 \\
23. R.Dijkgraaf  Intersection theory, integrable hierarchies and topological
field theory,

\hspace{.5in}
Preprint IASSNS$-HEP-91/91$, December 1991\\
24. M.Kontsevich  Intersection theory on the moduli space of curves and the
matrix Airy

\hspace{.5in}
function, Preprint $MPI/91-77$\\
25. M.Sato, Y.Sato  Soliton equations as dynamical systems in an
infinite-dimensional

\hspace{.5in}
Grassmannian, in: Nonlinear partial differential
equations in applied sciences,

\hspace{.5in}
P.Lax, H.Fujita, G.Strang (North-Holland,
Amsterdam, 1982)

E.Date, M.Jimbo, M.Kashiwara, T.Miwa  Transformation groups for soliton
equations,

\hspace{.5in}
RIMS Symp. ``Non-linear integrable systems -- classical theory and
quantum

\hspace{.5in}
theory (World scientific, Singapore, 1983)\\
26. M.Jimbo, T.Miwa, M.Sato  Holonomic quantum fields, $I-V:$ Publ.RIMS, Kyoto

\hspace{.5in}
Univ.,   {\bf 14} (1978) 223; {\bf 15} (1979) 201, 577, 871, 531\\
27. G.Segal, G.Wilson  Publ.Math. I.H.E.S., {\bf 61} (1985) 1

\end{document}